\documentclass[11pt,a4paper]{article}
\usepackage{jcappub}

\usepackage{amssymb}
\usepackage{amsfonts}
\usepackage{amsmath}
\usepackage{color}
\usepackage{graphicx}
\usepackage{rotating}
\usepackage{ifpdf}
\usepackage{overpic}
\usepackage{placeins} 

\title{Cosmic acceleration from modified gravity with Palatini formalism}
\author[a]{Andrzej Borowiec,}
\author[b]{Micha{\l} Kamionka,}
\author[c]{Aleksandra Kurek}
\author[c,d]{and Marek Szyd{\l}owski}
\affiliation[a]{Institute of Theoretical Physics, University of Wroc{\l}aw\\
pl. Maksa Borna 9, 50-204  Wroc{\l}aw, Poland.}
\affiliation[b]{Astronomical Institute, University of Wroc{\l}aw\\
ul. Kopernika 11, 51-622 Wroc{\l}aw, Poland.}
\affiliation[c]{Astronomical Observatory, Jagiellonian University\\
ul. Orla 171,
30-244 Krak{\'o}w, Poland.}
\affiliation[d]{Mark Kac Complex Systems Research Centre, Jagiellonian University\\
ul. Reymonta 4, 30-059 Krak{\'o}w, Poland.}
\emailAdd{borow@ift.uni.wroc.pl}
\emailAdd{kamionka@astro.uni.wroc.pl}
\emailAdd{kurek@oa.uj.edu.pl}
\emailAdd{uoszydlo@cyf-kr.edu.pl}
\abstract{We study new FRW type  cosmological models of modified gravity treated on the background of Palatini
approach.  These models are generalization of Einstein gravity by the presence of a scalar
field  non-minimally coupled to the curvature. The models employ Starobinsky's term in the
Lagrangian and dust matter. Therefore, as a by-product, an exhausted cosmological analysis of general relativity
amended by quadratic term is presented. We investigate dynamics of our models, confront them with the
currently available astrophysical data as well as against $\Lambda$CDM model.
We have used the dynamical system methods in order to investigate dynamics of the models.
It reveals the presence of a final sudden singularity. Fitting free parameters  we have demonstrated by statistical analysis that this class of models is in a very good agreement with the data (including CMB measurements)
as well as with the standard $\Lambda$CDM model predictions. One has to use statefinder diagnostic in order to discriminate among them. Therefore Bayesian methods of model selection have been employed
in order to indicate preferred model. Only in the light of CMB data the concordance model remains invincible.}
\arxivnumber{1109.3420}
\begin{document}
\maketitle

\newpage
\section{ Introduction}
As it is well-know a cosmological constant was the first and the simplest modification of general relativity performed by Einstein himself in order to stop cosmic expansion. Today the so-called Standard or Concordance (denoted also as $\Lambda$CDM or LCDM: Lambda Cold Dark Matter) Cosmological Model, which is based on this modification, turns out to be the best fitted model with respect to a huge amount of high precision currently available astrophysical data. In particular it properly describes a present day cosmic acceleration \cite{Perlmutter:1998np} by means of dark components: Dark Energy and Dark Matter. However $\Lambda$CDM model suffers for essential theoretical problems (e.g. well-known coincidence and fine tuning problems), specially related to a primordial stage of cosmic evolution \cite{reviews}. Some of these problems  can be cured by more sophisticated modifications. Among them  the so-called $f(R)$-gravity models constitute a huge family \cite{Sergei,f(R)rev} including also models based on Palatini formulation \cite{OlmoPalatRev}.
Quite recently there is a renewed interest in the Paltini modified gravity \cite{Palatini} which treats a metric and torsion-free connection as independent variables, see e.g. \cite{OlmoPalatRev,FFV,ABF} for more details. Resulting equations of motion remain second order as in the Einstein gravity case.

Modification of Einstein's General Relativity becomes a viable way to address
the accelerated expansion of the Universe as well as dark matter and dark energy problems in modern cosmology
(see e.g. \cite{reviews} and references therein). This includes modified  theories with
a non-trivial gravitational coupling \cite{DD3,ab1,Sergei2,Bertolami,Obukhov}. Viable non-minimal models unifying
early-time inflation with late-time acceleration have been, particularly, discussed in \cite{Sergei2}.
Apart of cosmological viability another justification for modified gravity should be taken seriously  into account: curvature corrections appear naturally as a low energy limits of quantum gravity, quantization on curved background and/or effects from extra dimensional physics.

Both purely metric as well as Palatini $f(R)$- gravity can be further extended by adding (scalar) field non-minimally
coupled to the curvature \cite{DD3,ab1,Bertolami}. In cosmological settings Palatini formalism gives rise, similarly
to Einstein gravity, to the first order autonomous differential equation for a scale factor which can be recast into the
form of Friedmann equation. This enables us to analyze the corresponding cosmological models as a one-dimensional
particle-like Hamiltonian system with entire dynamics encoded in an explicitly obtained effective potential function.
Such approach simplifies meaningly computer simulations and numerical analysis.
Before doing these one has to constrain model parameters by using astrophysical data \cite{union2,Simon2004,bao,Komatsu:2010fb} and CosmoNest
code \cite{Lewis1}. Simple Palatini based cosmological models has been previously tested against various sets of cosmological data in \cite{BGS}.
Very recently it has been shown that gravitational redshift of galaxies in clusters is consistent both with general relativity and some models of $f(R)-$modified gravity \cite{Wojtak}.

The paper is organized as follows. In Section 2 we recall some basic facts about Palatini gravity non-minimally coupled to dilaton-like field. Its cosmological application is considered in Section 3: particularly, general expression for generalized Friedmann equation is derived therein. Two classes of new cosmological models stemming from two different solutions of the so-called master equation are issued in Sections 4 and 5.
In Section 6 we describe estimation of models parameters by astrophysical data followed by Bayesian method of model selection provided in Section 7.
Dynamical analysis of our models by means of one-dimensional particle-like Hamiltonian system with a Newtonian type effective potential function is subject of our investigations in Section 8. In order to visualize the dynamics we watch plots of potential functions and draw two-dimensional phase space trajectories for the corresponding models. Section 9 treats about additional statefinder diagnostics. We end up with summary of obtained results and general conclusions presented in Section 10.

In this paper we took the first step in testing the kinematical sector of the cosmological models of modified gravity, i.e. we apply cosmographic analysis to the background dynamic. The next step will be the studying of matter perturbations in the class of models under consideration. In these future investigations the results of the present paper will be used as a starting point. This allows to exploit the advantages of Bayesian methods \cite{Trotta:2008qt}.

\section{ Preliminaries and notation}

The main object of our considerations in the present article is a cosmological application of some non-minimally coupled
scalar-tensor Lagrangians of the type \footnote{Throughout the paper we shall work with units $c=8\pi G=1$. The metric signature is $(- + + +).$ }
\begin{equation} \label{lag1}
L=\sqrt{g}\left( f(R)+F\left( R\right) L_{d}\right) +L_{mat}
\end{equation}
treated within the Palatini approach as in \cite{DD3}. Hereafter we set $L_{d}=-\frac{1}{2}g^{\mu \nu }\partial _{\mu }\phi \partial_{\nu }\phi$
the Lagrangian for a free-scalar (massless) dilaton-like field $\phi$ \footnote{One can easily add both mass and potential interaction for $\phi$, cf. \cite{DD3}. } and  $L_{\text{mat}}$ represents a matter Lagrangian.
Because of Palatini formalism  $R$ is a scalar  $R=R(g,\Gamma)=g^{\mu\nu}R_{\mu\nu}(\Gamma)$ composed of the metric $g$
and the Ricci tensor $R_{\mu\nu}(\Gamma)$ of the symmetric ($\equiv$ torsionless) connection $\Gamma$ (for more details concerning the Palatini formalism see e.g. \cite{FFV,ABF}).
Therefore $(g, \Gamma)$ are dynamical variables. Particularly, the metric $g$ is used for raising and lowering indices.

We began with recalling some general formulae already developed in \cite{DD3},
both  $f(R)$ and $F(R)$ are assumed to be analytical functions of $R$. Equations of motion for gravitational fields
$(\Gamma, g)$ can be recast \cite{DD3}
into the form of the generalized Einstein equations
\begin{equation} \label{hmeGEE4}
R_{\mu \nu }\left( h \right)\equiv  R_{\mu \nu }\left( bg\right)=g_{\mu \alpha }P_{\nu }^{\alpha }
\end{equation}
(see also \cite{ABF}), where
$R_{\mu \nu }(h)$ is now the Ricci tensor of the new conformally related metric $h=bg$ with the conformal factor
$b$ specified  below. A $(1,\,1)$ tensor $P_{\nu }^{\mu }$ is defined by
\begin{equation}\label{genE}
P_{\nu }^{\mu }=\frac{c}{2b} \delta _{\nu }^{\mu }-\frac{F\left(
R\right) }{b}T_{\ \ \ \nu }^{d\ \mu }+\frac{1}{b} T^{\text{mat}\ \mu} _{\ \ \ \ \ \nu }
\end{equation}
and contains
matter and dilaton stress-energy tensors:  $T^{\text{mat}} _{\mu\nu}=\frac{\delta L_{\text{mat}}}{\delta g_{\mu\nu}}$;
$T^{\text{d}} _{\mu\nu}=\frac{\delta L_{\text{d}}}{\delta g_{\mu\nu}}$.
Here one respectively has
\begin{equation}\label{bc}
\begin{cases}
c=\left( f\left( R\right) +F\left( R\right) L_{d}\right)
=(L-L_{mat})/\sqrt{g} \cr
b=f^{\prime }\left( R\right) +F^{\prime }\left( R\right) L_{d}
\end{cases}
\end{equation}
where a prime denotes the derivative with respect to $R$.

Field equations for the scalar  field $\phi$ is
\begin{equation}\label{dynd}
\partial _{\nu }\left( \sqrt{g}F(R)g^{\mu \nu }\partial _{\mu }\phi
\right) = 0
\end{equation}
Dynamics of the system (\ref{lag1}) is controlled by the so-called master equation
\begin{equation}
    bR=2c-F(R)L_d +\tau
\end{equation}
obtained by contraction of (\ref{hmeGEE4}), where we set $\tau =g^{\mu \nu }T _{\mu \nu }^{\text{mat}}$
for a trace of the matter stress-energy tensor. In more explicit form it reads as
\begin{equation}\label{master}
2f\left( R\right) -f^{\prime }\left( R\right) R+\tau =\left(
F^{\prime }\left( R\right) R-F\left( R\right) \right) L_{d}
\end{equation}
These  reproduce the same field equations as treated in \cite{DD3}.
\section{ Cosmology from the generalized Einstein
equations}
Assuming the Cosmological Principle to hold we take
the physical metric $g$ to be a  
Friedmann-Robertson-Walker (FRW) metric $g$
\begin{equation} \label{FRW}
g=-d t^2+a^2 (t) 
\Big[ {1 \over {1-\kappa r^2}} d r^2+ r^2 \Big( d \theta^2+\sin^2(\theta) d \varphi^2  \Big) \Big]
\end{equation}
where $a\equiv a (t)$ is a scale factor and $\kappa$ is the space curvature ($\kappa=0,1,-1$).
We also suppose that the matter content $T^{mat}_{\mu\nu}$ of the universe is
described by a non-interacting mixture of perfect fluids.
We denote by $w_i$ the corresponding barotropic coefficients. Each species is represented by the stress-energy tensor
$T^{(i)}_{\mu\nu}=(\rho_i+p_i)\,u_\mu u_\nu+p_i g_{\mu\nu}$ satisfying a metric (with the Christoffel connection of $g$) conservation equation $\nabla^{(g)\,\mu}T^{(i)}_{\mu\nu}=0$  (see \cite{Koivisto}). This gives rise to the standard relations between pressure and density
(equation of state) $p_i=w_i \rho_i$ and $\rho_i =\eta_i a^{-3\left(1+w_i\right) }$.

Let us recall that for the standard cosmological model based on the standard Einstein-Hilbert Lagrangian
(considered both in the purely metric as well as Palatini formalisms)
\begin{equation} \label{E-H}
L_{EH}=\sqrt{g}R+L_{mat}
\end{equation}
the corresponding Friedmann equation, ensuing from Einstein's field equations, takes the form
\begin{equation}\label{sfe}
H^2+\frac{\kappa}{a^2}\equiv \frac{\dot{a}^2}{a^2}+\frac{\kappa}{a^{2}}=
\frac{1}{3}\sum_i\eta_i\,a^{-3\left( 1+w_i\right) }
\end{equation}
when coupled to (non-interacting) multi-component non-interacting barotropic  perfect fluids, where
$\eta_i a^{-3\left( 1+w_i\right)}$ represents a perfect fluid component with an equation of state (EoS)
parameter $w_i$. Here $H={\dot{a}\over a}$ denotes a Hubble parameter.
This is due to the fact that geometry contributes to the r.h.s. of the Friedmann equation through (cf. \ref{master})
\begin{equation}\label{RScalar}
R=-\tau=\sum_i(1-3w_i)\eta_ia^{-3\left( 1+w_i\right) }\equiv \sum_i(1-3w_i)\rho_i
\end{equation}
For example, the preferred $\Lambda$CDM model can be defined by three fluid components: the cosmological constant $w_\Lambda=-1$, dust $w_{\text{dust}}=0$ and radiation $w_{\text{rad}}=\frac{1}{3}$ assuming the spacial flatness condition $\kappa=0$. (As a matter of fact the spacial curvature term $\kappa a^{-2}$ can be also mimicked by barotropic fluid $w_{curv}=-{1\over 3}$.) The radiation component which has no contribution to the trace $\tau$ can be  practically neglected due to extremely small value $\Omega_{rad}\sim 10^{-5}$. Alternatively, instead of introducing cosmological constant via perfect fluid Dark Energy component, one can, following Einstein, modify the Einstein-Hilbert Lagrangian (\ref{E-H}): $L_{EH}\rightarrow L_{\Lambda EH}=\sqrt{g}(R-2\Lambda)+L_{mat}$.

On the other hand we have that the field equation for the scalar field $\phi\equiv \phi(t)$ is
$ \frac{d}{dt}(\sqrt{g}F(R)\dot{\phi})=0$,
so that  $\sqrt{g}F(R)\dot{\phi }=\text{const}$ and consequently $gF(R)^{2}L_{d}=A ^{2}=\text{const}$.
This simply implies that
\begin{equation}
F(R)^{2}L_{d}=A ^{2}a^{-6} \label{tor}
\end{equation}
with an arbitrary positive integration constant $A^{2}$ (see (\ref{dynd})). It means that this term behaves as a stiff matter component ($w_{stiff}=1$).

Assuming perfect fluid matter as a source, the generalized Einstein equations rewrites under the form
\begin{equation}\label{genH}
\left( \frac{\dot{a}}{a}+\frac{\dot{b}}{2b}\right) ^{2}+\frac{\kappa}{a^{2}}=
\frac{F(R)L_d}{6b}\ +\frac{c}{6b}+\sum_i\frac{(1+3w_i)\eta_i}{6b}a^{-3\left( 1+w_i\right) }
\end{equation}
It becomes a generalized Friedmann equation for the ordinary Hubble parameter $H\equiv\frac{\dot{a}}{a}$ if we take into account that from
(\ref{master}) and (\ref{tor}) the scalars $R$ and $L_d$ are implicit functions of the scale factor $a$.
Further calculations give rise to the decomposition
\begin{equation}\label{genH2}
    H^2= K(a)G(a)
\end{equation}
where
 \begin{equation}\label{genH3}
    G(a)= \frac{f+2F L_d}{3} +\sum_i\frac{(1+3w_i)\eta_i}{3}a^{-3\left( 1+w_i\right) }- 2\kappa (f^\prime +F^\prime L_d)a^{-2}
\end{equation}
and the function $K(a)$ is defined by
\begin{equation}\label{ab1}
K(a)=2(f^\prime +F^\prime L_d)\left[2f^\prime -4F^\prime L_d +\frac{3[\tilde\tau +2(F^\prime R-F)L_d][f^{\prime\prime}+(F^{\prime\prime}-2F^{-1}(F^\prime)^2)L_d]}
{f^{\prime\prime}R-f^\prime+[F^{\prime\prime}R +2F^\prime -2F^{-1}(F^\prime)^2R]L_d}\right]^{-2}
\end{equation}
where $\tilde\tau=\sum_i(w_i+1)(1-3w_i)\eta_i\,a^{-3\left( 1+w_i\right) }$ for the multi-component fluid considered before: $\dot\tau=3H\tilde\tau$.
As explained above the parameters, $R, L_d$, and their functions, e.g. $f(R), F^\prime(R)L_d$, etc. become through equations
(\ref{master}), (\ref{tor}) implicitly dependent of the scale factor $a$. In this way, the generalized Friedmann provides
an autonomous system of first order ordinary differential equation for the scale factor $a$ (se the next Section for more details).
The decomposition (\ref{genH2}) is furnished in such a way that for standard cosmology: $b=1, c=R, L_d=0$ one has $K(a)=1$ and one recovers the standard Friedmann equation (\ref{sfe}), and particularly $\Lambda$CDM model as well.

Our objective here is to investigate a possible cosmological applications and confront against astrophysical data of the following subclass of gravitational Lagrangians (\ref{lag1})

\begin{equation}\label{lagr_ii2}
L=\sqrt{g}\left( R+\alpha R^2+\beta
R^{1+\delta}+\gamma R^{1+\sigma}L_{d}\right) +L_{mat}\end{equation}
where $\alpha, \beta, \gamma, \delta, \sigma$ are free parameters of the theory.  It should to be observed that the gravitational
part $f(R)$ contains the so-called Starobinsky quadratic term $R^2$ \cite{Starobinsky} with some $R^{1+\delta}$ contribution.
In the limit $\alpha\, ,\beta\, ,\gamma\rightarrow 0$ our Lagrangians reproduce General Relativity.
The constants $\alpha, \beta, \gamma$ are dimensionfull with the corresponding dimensions satisfying $[R]=[\alpha R^2]=..$ etc.. In fact, a numerical value of the constant $\gamma$ is unessential since it can be incorporated into the scalar field: $\phi\rightarrow |\gamma|^{1\over 2}\phi$. Therefore, we further assume that it takes a discreet values $\gamma=0, \pm 1$. Such re-scaling would not be possible if one admits nontrivial self-interaction with nonvanishing potential energy $U(\phi)$ for the scalar field.

Following a common strategy particularly applicable within the Palatini formalism (see \cite{FFV,ABF,ab1}) one firstly finds out an exact solution of the master equation (\ref{master}). It allows to construct explicit cosmological model based on the generalized Friedmann equation. For this purpose and in order to reduce a number of independent parameters, we shall assume  through the rest of this paper that universe is spatially flat ($\kappa=0$) and filled exclusively with the most natural dust matter component only:
$\tilde\tau=\rho= -\tau= \eta a^{-3}=2f-f^\prime R+(F-F^\prime R)L_d$. Modification comes from the geometric part of the theory.
In this case formulae (\ref{genH2})-(\ref{ab1}) simplify to more readable form
\begin{equation}\label{ab11}
H^2=\frac{2(f^\prime +F^\prime L_d)\left[3f-f^\prime R+(3F-F^\prime R)L_d\right]}{3\left[2f^\prime -4F^\prime L_d +\frac{3[2f-f^\prime R +(F^\prime R-F)L_d][f^{\prime\prime}+(F^{\prime\prime}-2F^{-1}(F^\prime)^2)L_d]}
{f^{\prime\prime}R-f^\prime+[F^{\prime\prime}R +2F^\prime -2F^{-1}(F^\prime)^2R]L_d}\right]^{2}}
\end{equation}
This reconstructs the $\Lambda$CDM model under the choice $f=R-2\Lambda,\ F=0$, which is the limit $\alpha=\gamma=0,\ \delta=-1,\  \beta=-2\Lambda$. Einstein-DeSitter universe arising from the Einstein--Hilbert action (\ref{E-H}) corresponds to $\alpha=\beta=\gamma=0$.

\section{ New cosmological models: solution I}

The first model we wish to be considered
is based on the solution of (\ref{master}), (\ref{tor}) resembling the Einstein--DeSitter universe (cf. \ref{RScalar})
\begin{equation}\label{standard}
R=\rho=\eta a^{-3}
\end{equation}
provided that the integration constant $A$ (see (\ref{tor})) and parameter $\sigma$ take the values
\begin{equation}
A^{2}=\gamma\frac{\beta(\delta-1)}{\delta}\eta^{2}  , \qquad \sigma =-\delta
\end{equation}
where $\delta\neq 0, 1$. Positivity of $A^2$ can be ensured by an appropriate choice of $\gamma=\pm 1$.
Remember that for the $\Lambda$CDM case one has $R=4\Lambda +\eta a^{-3}$ instead (\ref{standard}). The conformal factor $b$ reads now
\begin{equation}\label{ab2}
b=1+2\alpha\eta a^{-3}+\frac{3\delta-1}{\delta}\beta\eta^\delta a^{-3\delta}
\end{equation}
As a consequence, we have obtained the generalized Friedmann equations (\ref{genH2})-(\ref{ab1})
\[
H^2=K(a)G(a)
\]
under the form
\begin{eqnarray}\label{genFr2}
G(a)=\frac{2}{3}\eta a^{-3}+\frac{\alpha\eta^2}{3}a^{-6}-\frac{2-3\delta}{3\delta}\beta\eta^{\delta+1}a^{-3(\delta+1)}
\end{eqnarray}
and
\[
K(a)=\frac{2+4\alpha\eta a^{-3}-2\frac{1-3\delta}{\delta}\beta\eta^{\delta}a^{-3\delta}}{\left[2-2\alpha\eta a^{-3}-\frac{(1-3\delta)(2-3\delta)}{\delta}\beta\eta^{\delta}a^{-3\delta}\right]^2}.
\]
When $\alpha=0$ and $\beta=0$ the Friedmann equation takes the simplest form $H^2=\frac{1}{3}\eta a^{-3}$, i.e. the same as in Einstein general relativity with a pure dust matter. As we already mentioned it reconstructs the (flat) Einstein--DeSitter model.

Before proceeding further let us observe that the scaling properties of (\ref{genH}) are analogous to that in  standard cosmology (\ref{sfe}).
 It entitles us to introduce dimensionless cosmological (density like) parameters: the standard  $\Omega_{0,m}=\frac{\eta}{3H_0^2}$ as well the new one: $\Omega_{0,\beta}=\beta\eta^\delta$,
\,$\Omega_{0,\alpha}=\alpha\eta$ corresponding  to  new parameters $ \alpha$, $\beta$ in the Lagrangian (\ref{lagr_ii2}).
With this set of variables the generalized Friedmann equation rewrites under the form
\begin{eqnarray}\label{H_H0_2rozw}
\left(\frac{H}{H_0}\right)^2=K(z)G(z)=\frac{2+4\Omega_{0,\alpha}(1+z)^3-2\frac{1-3\delta}{\delta}\Omega_{0,\beta}(1+z)^{3\delta}}{\left[2-2\Omega_{0,\alpha}(1+z)^3-\frac{(1-3\delta)(2-3\delta)}{\delta}\Omega_{0,\beta}(1+z)^{3\delta}\right]^2}\times\cr
\times\left[2\Omega_{0,m}(1+z)^3+\Omega_{0,\alpha}\Omega_{0,m}(1+z)^6-\frac{2-3\delta}{\delta}\Omega_{0,\beta}\Omega_{0,m}(1+z)^{3(\delta+1)}\right]
\end{eqnarray}
where redshift $1+z=a^{-1}$.
The number of free parameters $(\alpha,\beta,\delta,\eta)=(\Omega_{0,\alpha}, \Omega_{0,\beta}, \delta, \Omega_{0,m})$ to be fitted by experimental data is 4. There are constrained by the following conditions: $\delta\neq0,-1,1$,  \hspace{0.4cm}$\Omega_{0,m}\in\langle0,1\rangle$\hspace{0.1cm}.

Above equation can be rewritten in a more convenient for computer simulation form
\begin{eqnarray}\label{H_H0_2rozw_i}
\left(\frac{H}{H_0}\right)^2=\Omega_{0,m}K(z)\tilde G(z)=\Omega_{0,m}\frac{2+4\Omega_{0,\alpha}(1+z)^3-2\frac{1-3\delta}{\delta}\Omega_{0,\beta}(1+z)^{3\delta}}{\left[2-2\Omega_{0,\alpha}(1+z)^3-\frac{(1-3\delta)(2-3\delta)}{\delta}\Omega_{0,\beta}(1+z)^{3\delta}\right]^2}\times\cr
\times\left[2(1+z)^3+\Omega_{0,\alpha}(1+z)^6-\frac{2-3\delta}{\delta}\Omega_{0,\beta}(1+z)^{3(\delta+1)}\right]
\end{eqnarray}

Now normalization constraint can be simply set by
\begin{equation}
\Omega_{0,m}K(0)\tilde G(0)=1.
\label{normalization}
\end{equation}
There is a number of interesting subcases, e.g, $\delta= {1\over 3}, {2\over 3}$ which we shall not study here. However the case of Einstein gravity supplemented by the quadratic Starobinsky term  corresponds to $\beta=0$: $L_{ES}=\sqrt{g}(R+\alpha R^2)+L_{mat}$. In this simplest case one gets
\begin{eqnarray}\label{EStarob}
\left(\frac{H}{H_0}\right)^2=
\Omega_{0,m}\frac{2+4\Omega_{0,\alpha}(1+z)^3}{\left[2-2\Omega_{0,\alpha}(1+z)^3\right]^2}\,\,
\left[2(1+z)^3+\Omega_{0,\alpha}(1+z)^6\right]
\end{eqnarray}
Cosmology for (\ref{EStarob}) has been also recently investigated in \cite{Koivisto2} (see also \cite{Starobinski2}).
We will see by comparison against experimental data that this model does not differ qualitatively from the $\Lambda$CDM, i.e. as expected it contains dynamical cosmological constant which becomes active in recent times. Adding dilaton field $\phi$ makes it even more quantitatively similar provided we shall use the same standard solution (\ref{standard}).

\section{ New cosmological models: solution II}

Yet another models can be determined by the  solution
\begin{equation}\label{news}
R=
\xi a^{-\frac{3 }{1+\delta}} , \qquad A^{2}=\frac{\gamma}{2\delta}\left[\frac{\eta}{\left(
1-\delta\right)\beta}\right]^{2} , \qquad \sigma=2\delta
\end{equation}
where $\xi\equiv\left[\frac{\eta }{\left( 1-\delta\right)\beta }\right]^{\frac{1}{1+\delta}}$.
 One needs  $\beta(1-\delta),\,\gamma\delta>0$. Now the parameter $\beta$ cannot vanish.
The conformal factor $b$ reads
\begin{equation}\label{ab}
b
=\frac{1+4\delta}{2\delta}+2\alpha \xi a^{-\frac{3}{1+\delta}}+\beta (1+\delta) \xi^\delta a^{-\frac{3\delta}{1+\delta}}
\end{equation}
As a consequence, we obtain components constituting the generalized Friedmann equations  (\ref{genH2})-(\ref{ab1}) (cf. (\ref{ab11})) under the form
\begin{eqnarray}\label{genFr}
 G(a)= \frac{1+\delta}{3\delta}
\xi a^{-\frac{3}{1+\delta}}\ +\frac{1}{3}\alpha\xi^2 a^{-\frac{6}{1+\delta}}\ +\frac{2-\delta}{3(1-\delta)}\eta a^{-3} \cr
K(a)=\frac{\frac{1+4\delta}{\delta}+4\alpha\xi a^{-\frac{3}{1+\delta}}
+2\beta(1+\delta)\xi^{\delta} a^{-\frac{3\delta}{1+\delta}}}
{\left[\frac{1+4\delta}{\delta}+2\frac{2\delta-1}{1+\delta}\alpha\xi a^{-\frac{3}{1+\delta}}
+\beta(2-\delta)\xi^{\delta}a^{-\frac{3\delta}{1+\delta}}\right]^2}
\end{eqnarray}

Introducing, as before,  dimensionless density parameters: $\Omega_{0,m}=\frac{\eta}{3H_0^2}$,  $\Omega_{0,\beta}=\frac{\xi}{3H_0^2} =\frac{1}{3H_0^2}\left[\frac{\eta}{(1-\delta)\beta}\right]^{\frac{1}{1+\delta}}$,
\,$\Omega_{0,\alpha}=\alpha H_0^2\Omega_{0,\beta}$ one gets in terms of the redshift $1+z=a^{-1}$
\begin{eqnarray}\label{H_H0}
\left(\frac{H}{H_0}\right)^2=K(z)G(z)=\frac{\frac{1+4\delta}{\delta}+12\Omega_{0,\alpha}(1+z)^{\frac{3}{1+\delta}}
+2\frac{1+\delta}{1-\delta}\Omega_{0,m}\Omega_{0,\beta}^{-1}(1+z)^{\frac{3\delta}{1+\delta}}}
{\left[\frac{1+4\delta}{\delta}+6\frac{2\delta-1}{1+\delta}\Omega_{0,\alpha}(1+z)^{\frac{3}{1+\delta}}
+\frac{2-\delta}{1-\delta}\Omega_{0,m}\Omega_{0,\beta}^{-1}(1+z)^{\frac{3\delta}{1+\delta}}\right]^2}\times\cr
\times\left[\frac{1+\delta}{\delta}\Omega_{0,\beta}(1+z)^{\frac{3}{1+\delta}}
+3\Omega_{0,\alpha}\Omega_{0,\beta}(1+z)^{\frac{6}{1+\delta}}+\frac{2-\delta}{1-\delta}\Omega_{0,m}(1+z)^3\right]
\end{eqnarray}
The number of free parameters $(\alpha,\beta,\delta,\eta)=(\Omega_{0,\alpha}, \Omega_{0,\beta}, \delta, \Omega_{0,m})$ to be fitted by experimental data is 4. There are constrained by the following conditions: $(1-\delta)\beta, \eta, \Omega_{0,\beta} >0$,\hspace{0.4cm}$\Omega_{0,m}\in\langle0,1\rangle$,\hspace{0.4cm}
$\delta\neq0,-1,1$.

Introducing $\Omega_{0,c}=\Omega_{0,m}\Omega_{0,\beta}^{-1}$ allows us to rewrite the last equation in a more convenient form
\begin{eqnarray}\label{H_H0_omegac}
\left(\frac{H}{H_0}\right)^2=\Omega_{0,\beta}K(z)\tilde G(z)=\Omega_{0,\beta}\frac{\frac{1+4\delta}{\delta}+12\Omega_{0,\alpha}(1+z)^{\frac{3}{1+\delta}}
+2\frac{1+\delta}{1-\delta}\Omega_{0,c}(1+z)^{\frac{3\delta}{1+\delta}}}
{\left[\frac{1+4\delta}{\delta}+6\frac{2\delta-1}{1+\delta}\Omega_{0,\alpha}(1+z)^{\frac{3}{1+\delta}}
+\frac{2-\delta}{1-\delta}\Omega_{0,c}(1+z)^{\frac{3\delta}{1+\delta}}\right]^2}\times\cr
\times\left[\frac{1+\delta}{\delta}(1+z)^{\frac{3}{1+\delta}}
+3\Omega_{0,\alpha}(1+z)^{\frac{6}{1+\delta}}+\frac{2-\delta}{1-\delta}\Omega_{0,c}(1+z)^3\right]
\end{eqnarray}
Then normalization constraint can be simply set by
\begin{equation}
\Omega_{0,\beta}K(0)\tilde G(0)=1.
\label{normalization_ii}
\end{equation}
Again in order  to better control the role of Starobinsky term one can switch it off by setting $\alpha=0$. As it has
been mentioned before the parameter $\beta$ cannot vanish now.

One can summarize this part by concluding that we have obtained for numerical analysis four new cosmological  models
which will be further on denoted correspondingly as $I, I_{\alpha=0}$ and $II, II_{\alpha=0}$.
Models $I$ and $II$ have three free parameters to be fitted by experimental data. Models $I_{\alpha=0}$ and
$II_{\alpha=0}$ have two such parameters. These models are formulated not only in terms of Lagrangian functions
but also by giving explicit form for the corresponding Friedmann equations.
Finally, the simplest case $I_{\beta=0}$ describes  
cosmological model  based on $R^2$ Palatini modified gravity without dilaton field which has similarly to
$\Lambda$CDM only one free parameter. Such quadratic gravity models have been extensively studded in the literature (\ref{EStarob})
by using different methods than the one we have employed here. The reason is that the explicit form of the corresponding
Friedmann equation for this case (see (\ref{EStarob})) has not been used except \cite{Koivisto2}.

\section{ Constraining model parameters by astrophysical data}

In the Bayesian approach to estimation model parameters (i.e. best fit values and credible intervals) one uses a posterior probability density function (pdf), which is defined in the following way:
\begin{equation}\label{posterior}
 P(\bar{\Theta}|D,M)=\frac{P(D|\bar{\Theta} , M) P(\bar{\Theta} |M)}{P(D|M)}.
\end{equation}
$\bar{\Theta}$ is the vector of model $M$ parameters and $P(D|\bar{\Theta} , M) \equiv L $ is the likelihood function for model $M$. The so called prior pdf for model parameters $P(\bar{\Theta} |M)$ should be estimated before the data $D$ comes into analysis and should involve information which we have gathered in earlier studies, e.g. with different data sets or on theoretical grounds. Finally $P(D|M) \equiv E$ is the
so-called evidence, which could be ignored in model constraint analysis. The best fit values can be estimated using the maximum of the joined posterior pdf (\ref{posterior}) (i.e. its mode). One can also consider marginalized posterior pdf:
\begin{equation}\label{marg_post}
P(\theta_i | D,M) = \int  P(\bar{\Theta}|D,M) \mathrm{d}  \bar{\phi},
\end{equation}
where $\bar{\Theta}=\{\theta_i,\bar{\phi}$\}, and use its mode or mean and credible interval (usually defined as interval which involves $68\%$ or $95\%$ of the probability).

In order to estimate the parameters of our models we use supernovae (SNIa) data \cite{union2}, the observational $H(z)$ data \cite{Simon2004}, the measurement of the baryon acoustic oscillations (BAO) from the SDSS luminous red galaxies \cite{bao} and information coming from CMB \cite{Komatsu:2010fb}.

We use a sample of $N=557$ data \cite{union2}, which consist of data from Union \cite{Kowalski:2008ez}, Supernova Cosmology Project \cite{union2}, SDSS SN Survey \cite{Holtzman:2008zz} and CfA3 \cite{Hicken:2009dk}. The likelihood function is defined in the following way:
\begin{equation}
L_{SN} \propto \exp \left[ - \sum_{i}\frac{(\mu_{i}^{\mathrm{theor}}-\mu_{i}^{\mathrm{obs}})^{2}}
{2 \sigma_{i}^{2}} \right]  ,\label{chi2}
\end{equation}
where: $\sigma_{i}$ is the total measurement error, $\mu_{i}^{\text{obs}}=m_{i}-M$ is the measured value ($m_{i}$--apparent magnitude, $M$--absolute magnitude of SNIa), $\mu_{i}^{theor}=5\log_{10}D_{Li} + \mathcal{M}=5\log_{10}d_{Li} + 25$, $\mathcal{M}=-5\log_{10}H_{0}+25$ and $D_{Li}=H_{0}d_{Li}$, where $d_{Li}$ is the luminosity distance given by $d_{Li}=(1+z_{i})c\int_{0}^{z_{i}} \frac{dz'}{H(z')}$ (with the assumption $k=0$). In this paper the likelihood as a function independent of $H_0$ has been used (which is obtained after analytical marginalization of formula (\ref{chi2}) over $H_0$).

We use constraints coming from $13$ measurements of Hubble function at different redshifts $z$. Those data points are obtained by two observational methods. The first method is based on the measurements of spectroscopic ages of red galaxies \cite{Stern:2009ep}, while the second one is based on the measurements of BAO scale in radial direction \cite{Gaztanaga:2008xz}. For the $H(z)$ data the likelihood function is given by:
\begin{displaymath}
L_{H_z} \propto \exp \left[ - \sum_i\frac{\left(H(z_i)-H_i\right)^2}{2 \sigma_i^2} \right ],
\end{displaymath}
where $H(z_i)$ is the Hubble function, $H_i$ denotes observational data.

We also use information coming from the so called BAO A parameter, which is related to the Baryon Acoustic Oscillations scale measured in the redshift space power spectrum of luminous red galaxies (LRG) from the Sloan Digital Sky Survey (SDSS) \cite{bao}. For BAO A parameter data the likelihood function is characterized by:
\begin{equation}
L_{BAO} \propto \exp \left[  -\frac{(A^{\text{theor}}-A^{\text{obs}})^2}{2\sigma_A^2} \right] ,
\end{equation}
where $A^{\text{theor}}=\sqrt{\Omega_{\text{m},0}} \left (\frac{H(z_A)}{H_{0}} \right ) ^{-\frac{1}{3}} \left [ \frac{1}{z_{A}} \int_{0}^{z_{A}}\frac {H_0}{H(z)} dz\right]^{\frac{2}{3}}$ and $A^{\text{obs}}=0.469 \pm 0.017$ for $z_{A}=0.35$.

Finally, we use constraints coming from CMB temperature power spectrum, ie. CMB $R$ shift parameter \cite{Bond:1997}, which is related to the angular diameter distance ($D_A (z_*)$) to the last scattering surface:
\begin{equation}
R=\frac{\sqrt{\Omega_m H_0}}{c}(1+z_*)D_A(z_*).
\end{equation}
The likelihood function has the following form:
\begin{equation}
L_{CMB} \propto \exp \left[ -\frac{1}{2}\frac{(R-R_{obs})^2}{\sigma_{A}^2} \right],
\end{equation}
where $R_{obs}=1.725$ and $\sigma_{A} ^{-2}=6825.27$ for $z_*=1091.3$ \cite{Komatsu:2010fb}.

The entire $L_{TOT}$ is characterized by:
\begin{equation}
L_{TOT}=L_{SN}L_{H_z}L_{BAO}L_{CMB}.
\label{total_chi2}
\end{equation}
We assume flat prior probabilities for model parameters (second column of Table \ref{zestawienie})~. The prior probabilities because of constraints coming from previous estimations of the parameters were calculated using $N=192$ SNIa data sample \cite{Davis:2007na}. Additionally we have assumed that $H_0=74.2 \ [kms^{-1}Mpc^{-1}]$  \cite{0004-637X-699-1-539} and $\Omega_m \in [0,1] $.

The knowledge of prior pdf for model parameters and the formula for likelihood function enable us to calculate joined and marginalized posterior pdf. The mode of joined posterior pdf as well as mean (together with $68\%$ credible interval) of marginalized posterior pdf were calculated using publicly available  CosmoNest package \cite{Lewis1,Mukherjee:2005wg,Mukherjee:2005tr,Parkinson:2006ku}, which was modified for our purpose.

In table~\ref{zestawienie} we display the best fitted parameters for all our models $I_{\alpha=0},\ I_{\beta=0},\ I,\ II_{\alpha=0},\ II$ as well as for $\Lambda$CDM estimated by  CosmoNest package. Parameters $\Omega_{0,\beta}$, $\Omega_{0,m}$ for models I and parameter $\Omega_{0,m}$ for models II are calculated using equations (\ref{normalization}), (\ref{normalization_ii}) and estimated values of the remaining parameters. We consider two cases: estimations with data sets coming from late universe (i.e. SNIa, H(z) and BAO) and estimations including also information from early universe (i.e. SNIa, H(z), BAO and CMB). Top part of the table relates to the first case, while the bottom relates to the second one. The best fit values correspond to the mean of the marginalized posterior pdf ($68\%$ credible intervals are also shown, estimation of sample variance is shown in cases where the credible interval has misleading values). Finally mode of joined posterior pdf is given (values in brackets) and the corresponding value of $\chi^2$ (table \ref{evidence}) (see next Section for explanations).

\begin{table}
\begin{center}
\tiny\begin{tabular}{|c|c|c||c|c||c|}
  \hline
   &$\Omega_{0,c}$&$\delta$&$\Omega_{0,\beta}$&$\Omega_{0,m}$&$\chi^2_{TOT}/2$\\
  \hline
  $\Omega_{0,\alpha}=0$,\, $\Omega_{0,c} \in <-1,10>$,\, $\delta \in (0,1>$&
  $2.572$&$0.997$&$0.254$&$0.652$&$294.233$\\
  $\Omega_{0,\alpha}=-100$,\, $\Omega_{0,c} \in <-1,10>$,\, $\delta \in (0,1>$& $1.748$ & $0.553$ & $0.012$& $0.045$&$278.974$\\
  $\Omega_{0,\alpha}=-150$,\, $\Omega_{0,c} \in <-1,10>$,\, $\delta \in (0,1>$& $2.226$ & $0.538$ & $0.005$& $0.024$&$278.853$\\
  $\Omega_{0,\alpha}=-300$,\, $\Omega_{0,c} \in <-1,10>$,\, $\delta \in (0,1>$& $3.469$ & $0.524$ & $0.009$& $0.013$&$278.737$\\
  $\Omega_{0,\alpha}=-1000$,\, $\Omega_{0,c} \in <-1,10>$,\, $\delta \in (0,1>$& $4.693$ & $0.508$ & $0.020$& $0.019$&$278.655$\\
  \hline
\end{tabular}
\caption{Comparison of estimated parameters $\Omega_{0,c}$, $\delta$, $\Omega_{0,\beta}$ and $\Omega_{0,m}$ for model II with fixed value of $\Omega_{0,\alpha}$. It shows that essential parameters as $\delta$ or $\Omega_{0,m}$ behaves stable under a wide range of $\Omega_{0,\alpha}$ provided $\Omega_{0,\alpha}\neq 0$.}
\label{fixed_oa}
\end{center}
\end{table}

In order to get better insight into the role of quadratic term we have made several
estimations for fixed values of $\Omega_\alpha$. The results are gathered in table
\ref{fixed_oa}. They differ significantly when $\alpha=0$ and $\alpha\neq 0$. In the last case the concrete values
of $\Omega_\alpha$ have secondary meaning. Moreover, the switching off  $R^2$ term leads to the cosmology which is not longer similar to $\Lambda$CDM (cf. Fig \ref{potencjal_oa_0}, \ref{potencjal_oa_0_ii}).

\begin{figure}[h!!!!]
\centering
   \includegraphics[width=0.75\textwidth]{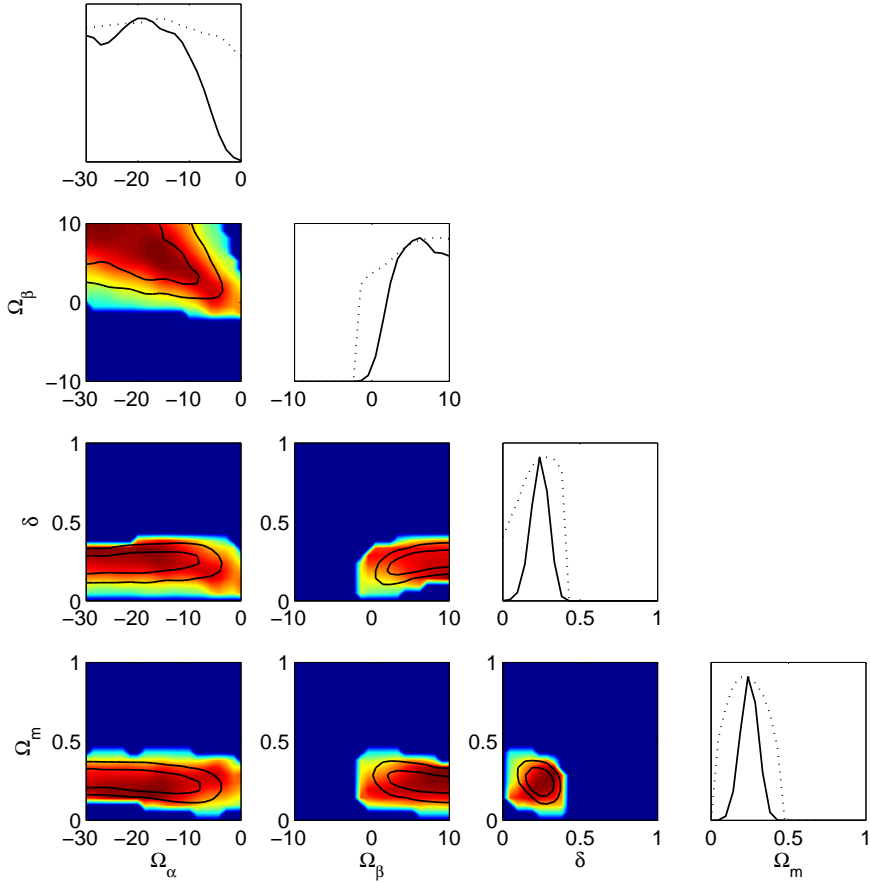}\\
   \caption{Constraints of the parameters of model $I$ -- estimations without CMB data. In 2D plots solid lines are
   the 68\% and 95\% confidence intervals from the marginalized probabilities. The colors reflect the mean likelihood of
   the sample. In 1D plots solid lines show marginalized probabilities of the sample, dotted lines are mean likelihood.
   For numerical results see Table \ref{zestawienie}, No 3.}
   \label{nowy_wzor_ii}
\end{figure}

\begin{figure}[h!!!!]
\centering
   \includegraphics[width=0.75\textwidth]{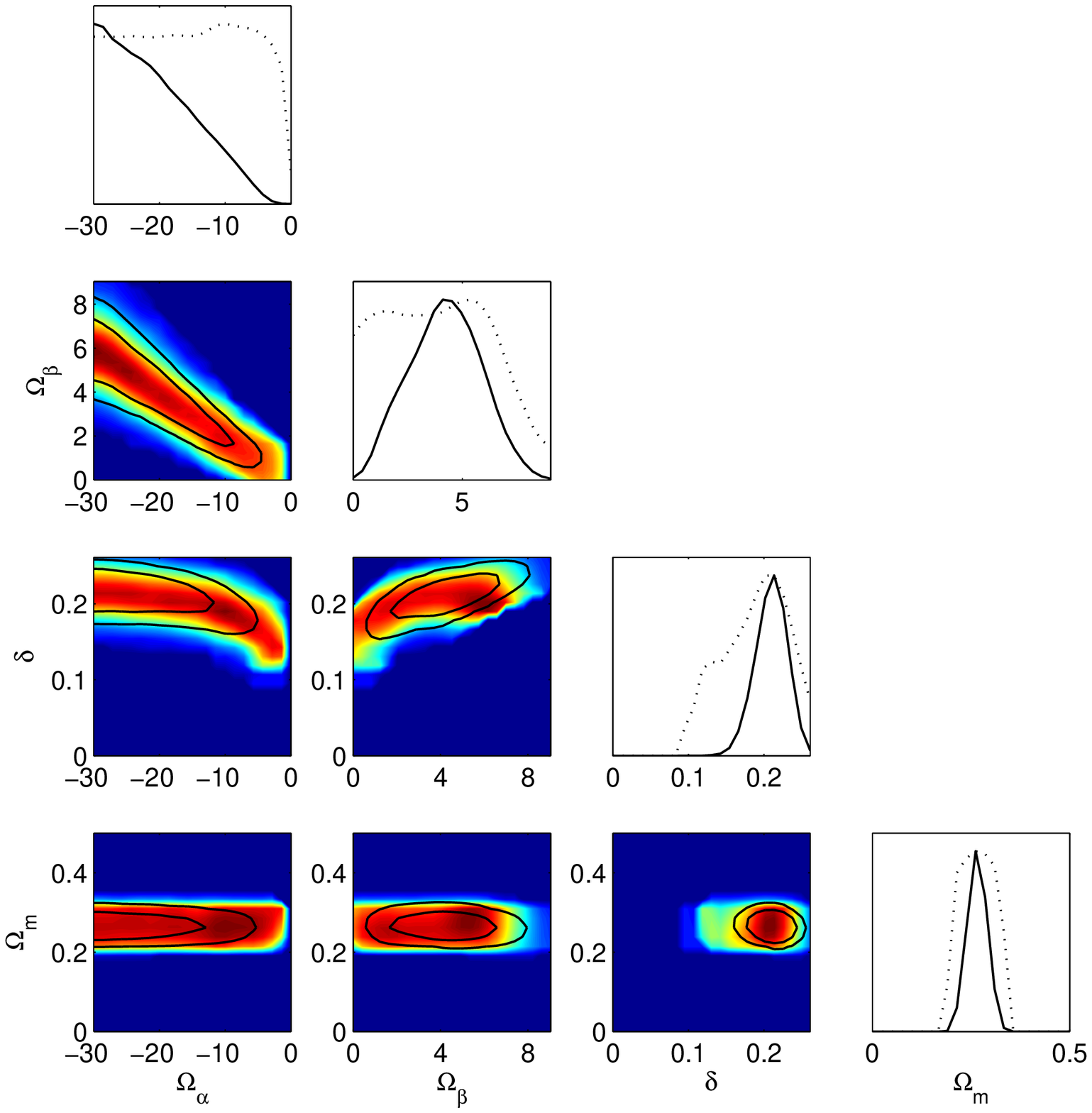}\\
   \caption{Constraints of the parameters of model $I$ -- estimations including CMB data.
   In 2D plots solid lines are the 68\% and 95\% confidence intervals from the marginalized probabilities.
   The colors reflect the mean likelihood of the sample. In 1D plots solid lines show marginalized probabilities of
   the sample, dotted lines are mean likelihood. For numerical results see Table \ref{zestawienie}, No 9.}
   \label{cmb_nowy_wzor_tri}
\end{figure}

\begin{figure}[h!!!!]
\centering
   \includegraphics[width=0.75\textwidth]{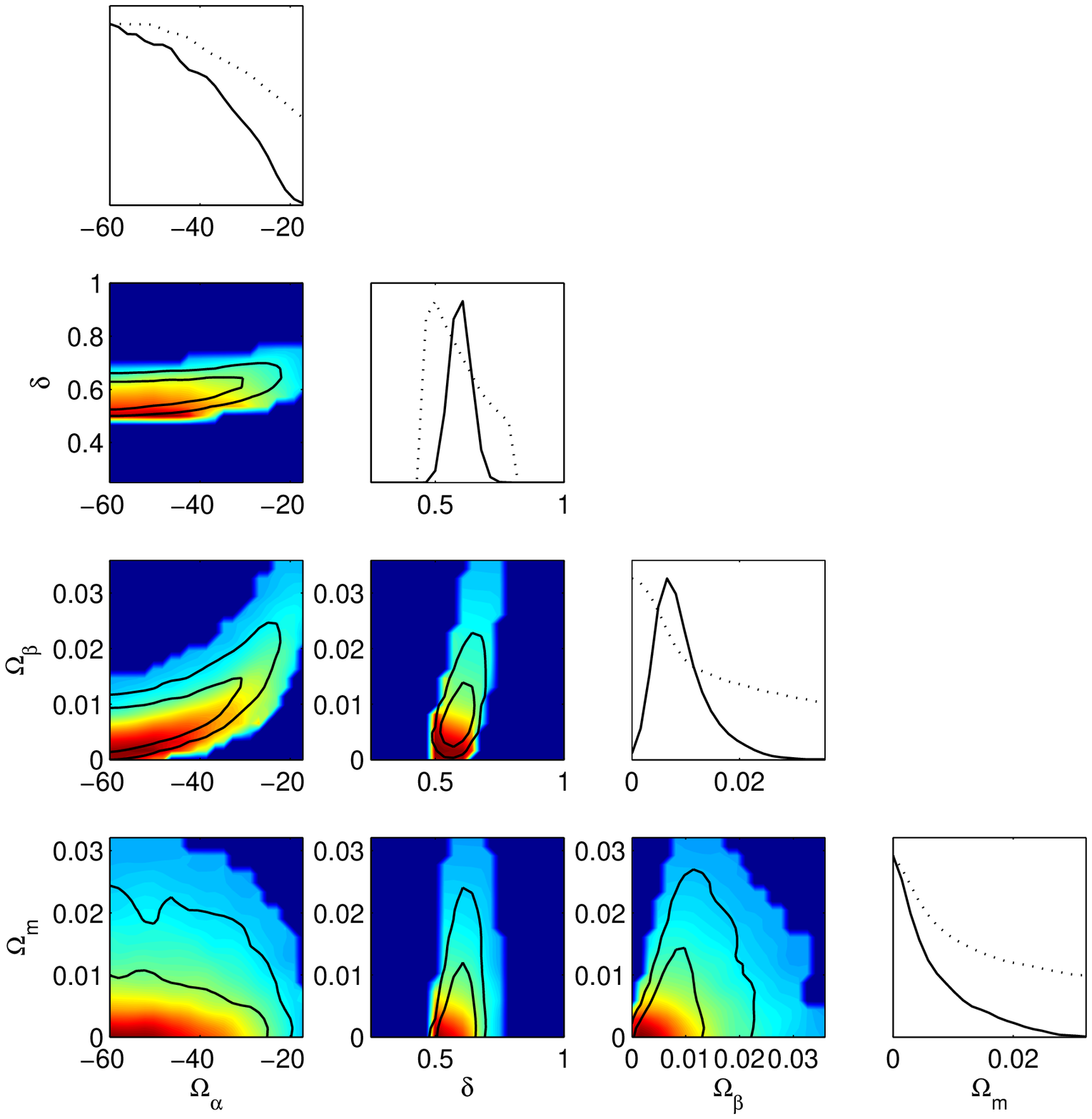}\\
   \caption{Constraints of the parameters of model $II$ -- estimations without CMB data. In 2D plots solid lines are
   the 68\% and 95\% confidence intervals from the marginalized probabilities. The colors reflect the mean likelihood
   of the sample. In 1D plots solid lines show marginalized probabilities of the sample, dotted lines are mean likelihood.
   For numerical results see Table \ref{zestawienie}, No 5.}
   \label{nowe_dane_tri}
\end{figure}

\begin{figure}[h!!!!]
\centering
   \includegraphics[width=0.75\textwidth]{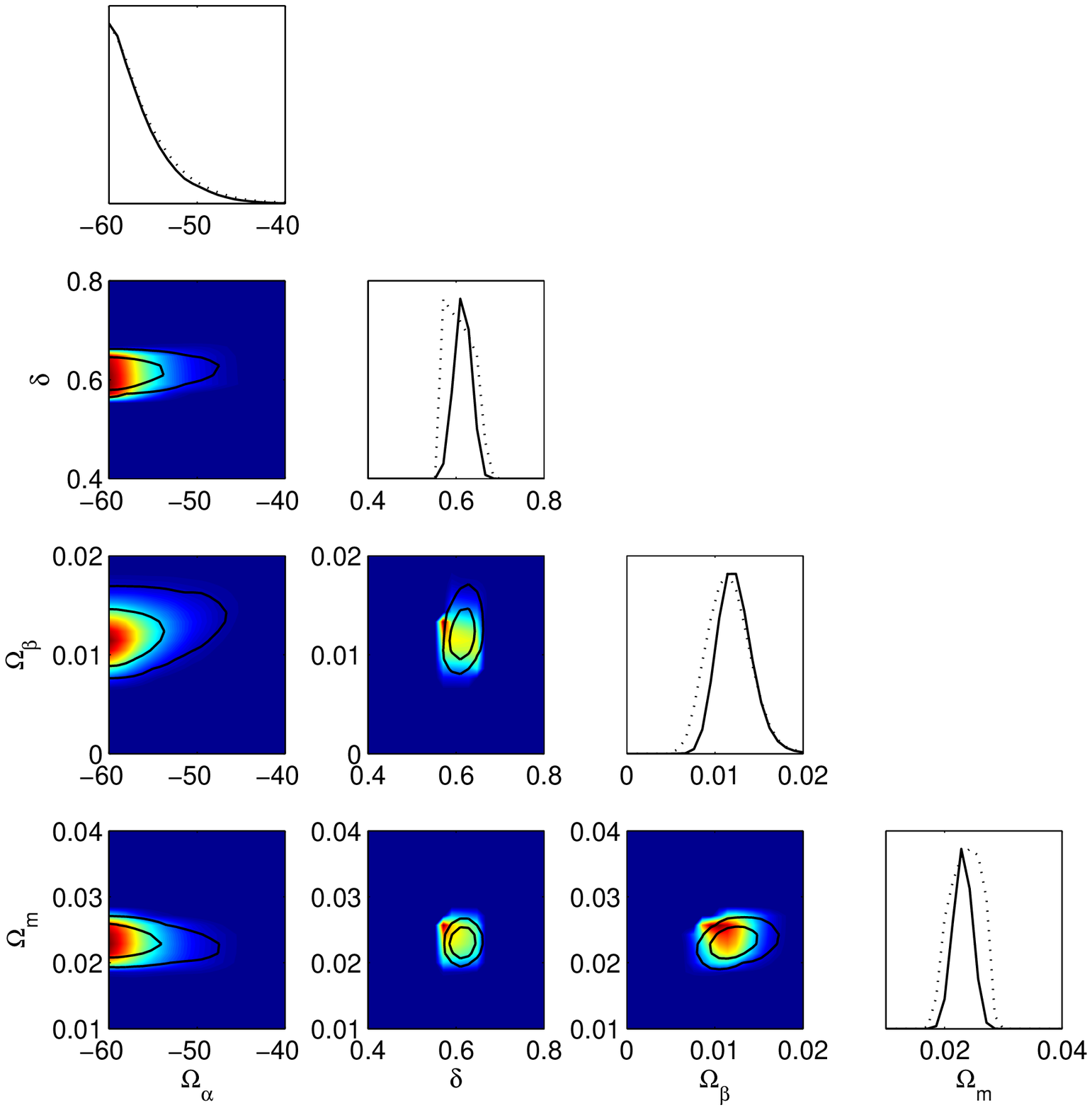}\\
   \caption{Constraints of the parameters of model $II$ -- estimations including CMB data.
   In 2D plots solid lines are the 68\% and 95\% confidence intervals from the marginalized probabilities.
   The colors reflect the mean likelihood of the sample. In 1D plots solid lines show marginalized probabilities
   of the sample, dotted lines are mean likelihood. For numerical results see Table \ref{zestawienie}, No 11.}
   \label{cmb_nowe_dane_tri}
\end{figure}

At the fig. \ref{nowy_wzor_ii}, \ref{nowe_dane_tri}, \ref{cmb_nowy_wzor_tri}, \ref{cmb_nowe_dane_tri} we show
constraints of the parameters of our models. In the 2D plots solid lines are the 68\% and 95\% confidence intervals
from the marginalized probabilities. The colors reflect the mean likelihood of the sample. In the 1D plots solid lines
show marginalized probabilities of the sample, dotted lines are mean likelihood. It is seen that it's not possible
to constrain properly parameter $\Omega_{0,\alpha}$ using available data (cf. table \ref{fixed_oa}).

\begin{figure}[h!!!!]
\centering
   \includegraphics[width=0.496\textwidth]{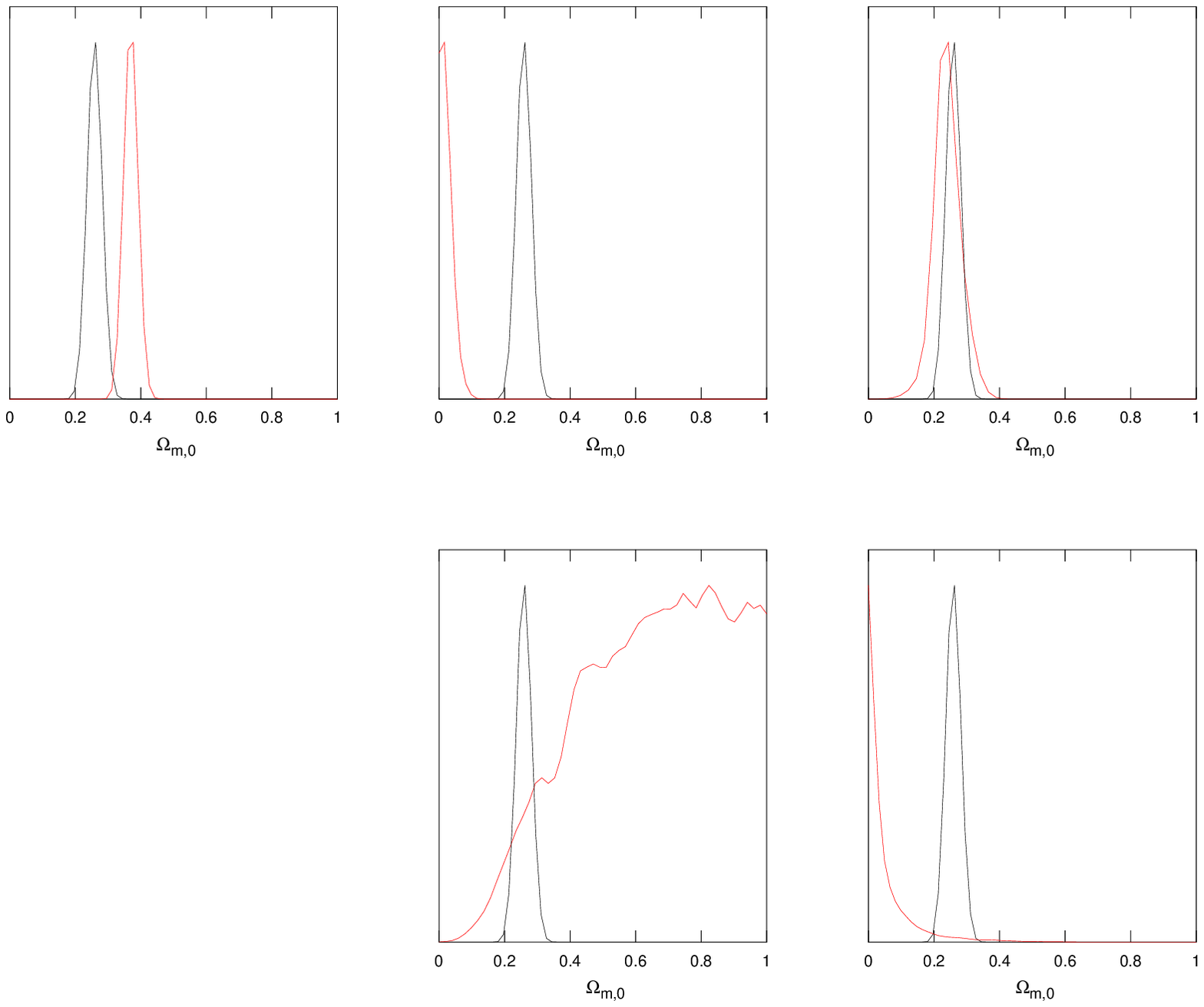}
   \includegraphics[width=0.496\textwidth]{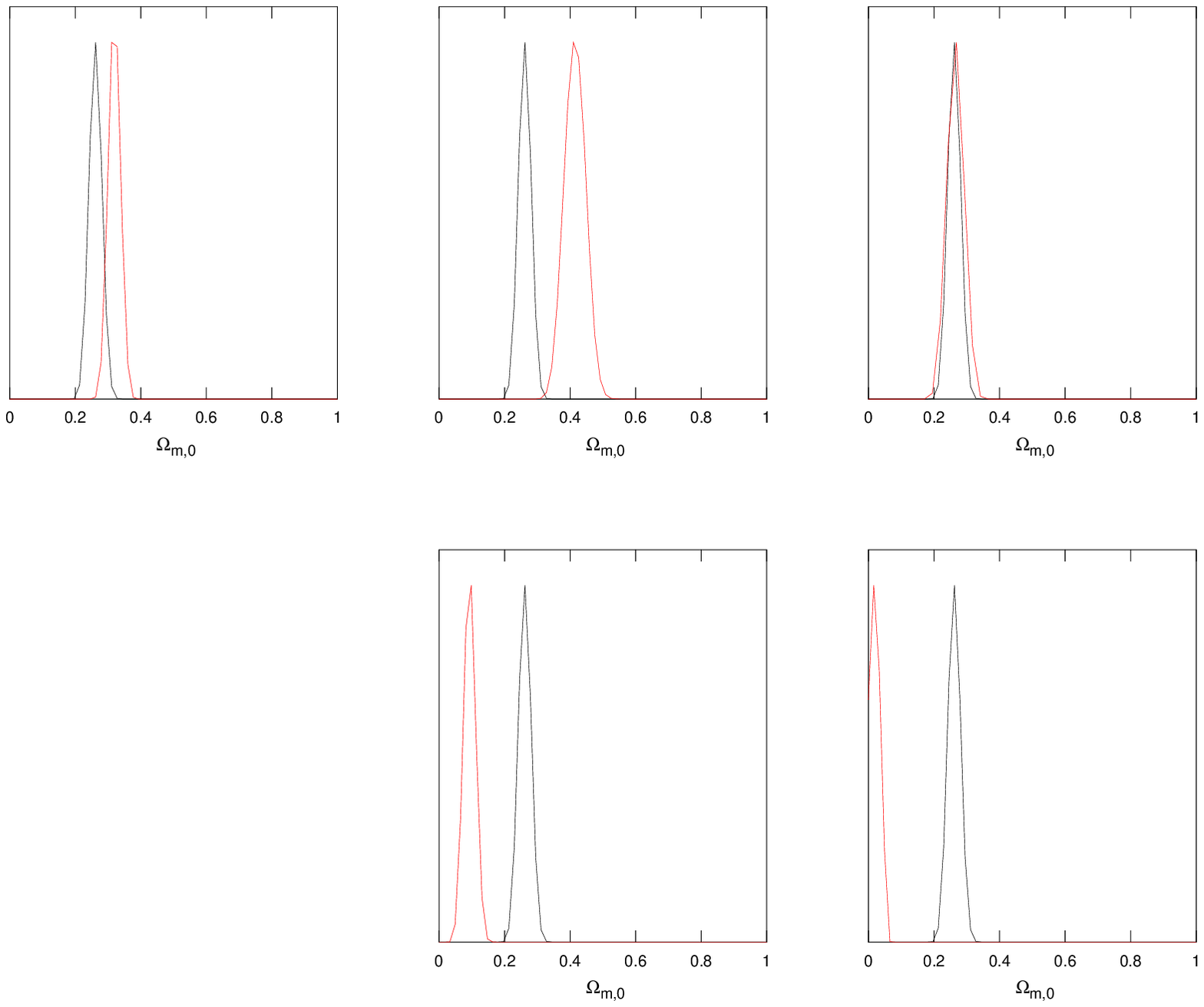}\\
   \caption{Posterior probability density functions of $\Omega_{m,0}$ parameter for all cases (red lines).
   First row correspond to models I, second to models II.
   First column and forth: $\beta=0$ (i.e. quadratic gravity); second and fifth: $\alpha=0$ .
    In third and sixth column 3 parameters were fitted. Black curves correspond to $\Lambda$CDM model. Left panel: parameters were fitted using SnIa, $H_z$ and BAO data.
   Right panel includes CMB data.  In some cases $\Omega_{m,0}$ is the same order as $\Omega_{barionic}\sim 0.04$.
   For numerical values see Table \ref{zestawienie}.}
   \label{omegaM}
\end{figure}

It is interesting to compare (Fig~\ref{omegaM}) a posterior probability density functions of $\Omega_{0, m}$ for new models with the one for $\Lambda$CDM model  (see  table~\ref{zestawienie} for the corresponding best fit values and credible intervals).
As one can conclude only for model $I$ (all three parameters fitted with and without CMB data) it remains at
the same level when compared with $\Lambda$CDM model. It should be remarked that in some cases, best fit value for $\Omega_{0,m}$ is small and close to the well-known amount of barionic matter $\Omega_b\sim 0,05$ (cf. table~\ref{zestawienie} row no 1, 5, 10, 11).

\newpage
\begin{figure}[h!!!!]
\centering
   \includegraphics[width=0.347\textwidth,angle=-90]{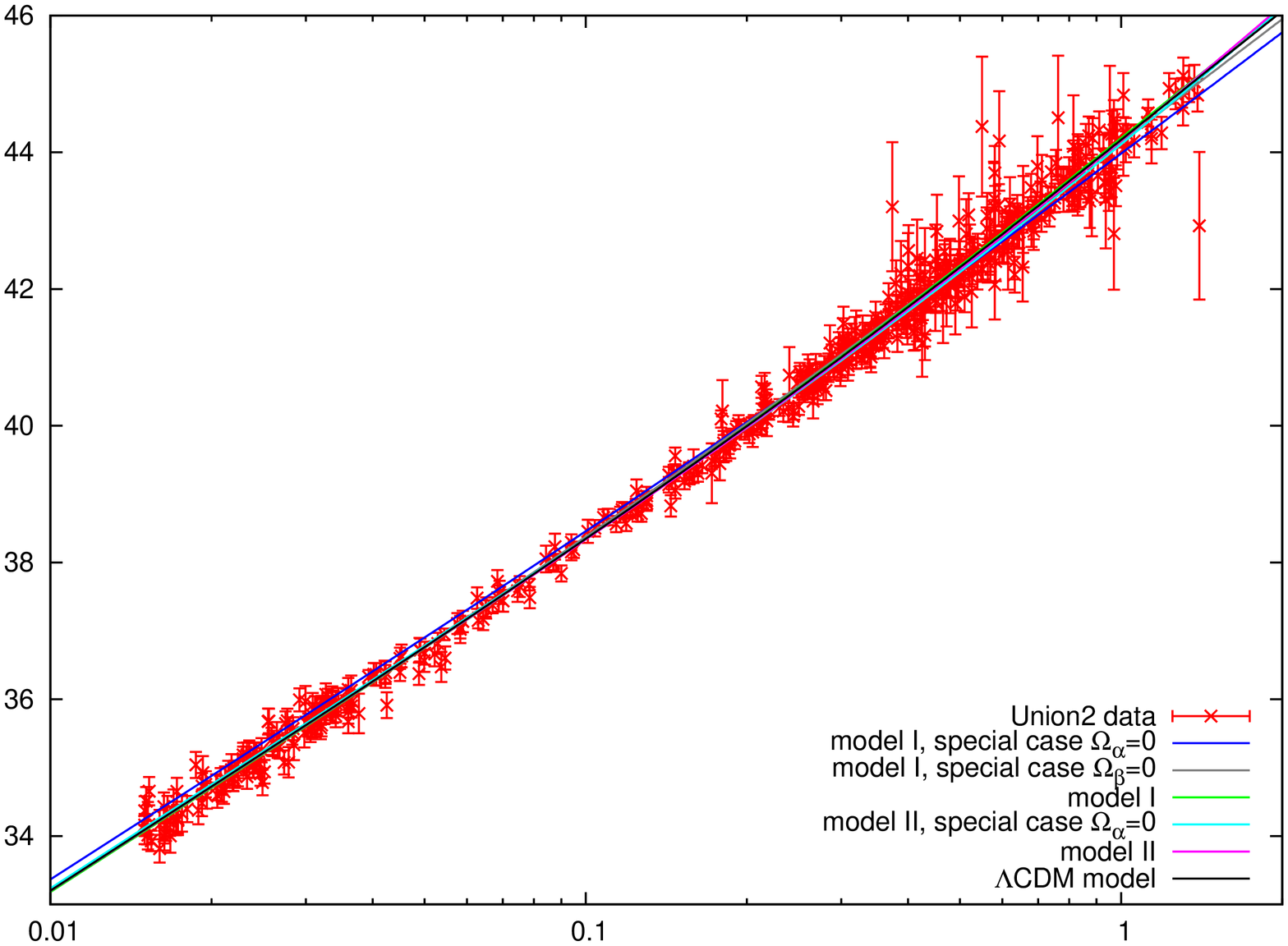}
   \includegraphics[width=0.347\textwidth,angle=-90]{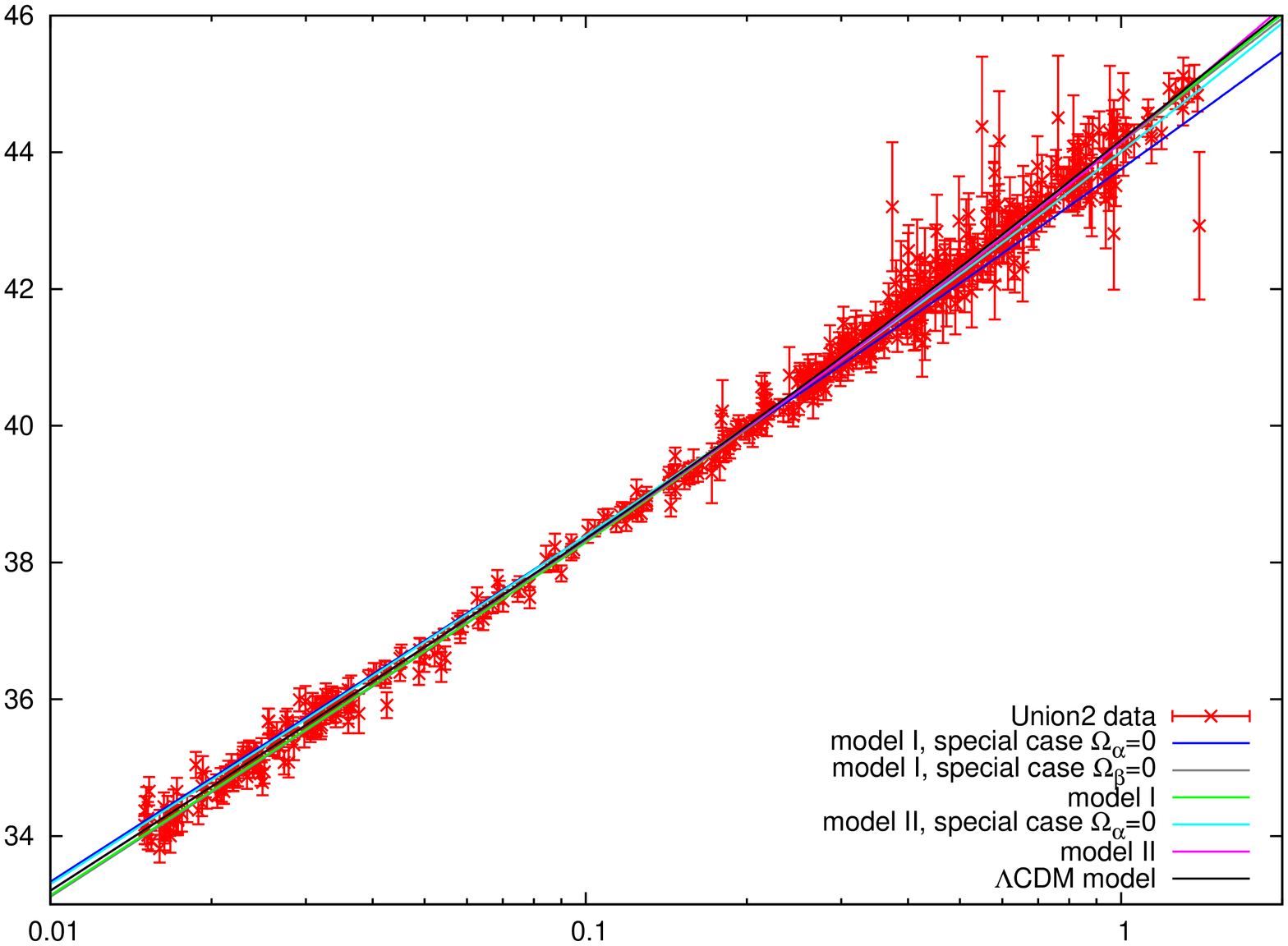}\\
   \caption{Comparison of Hubble's diagrams for model I (green) and II (magenta). Grey line denotes special case of
   quadratic gravity $I_{\beta=0}$. Blue ($I_{\alpha=0}$) and light blue ($II_{\alpha=0}$) lines denote most divergent
   with respect to $\Lambda$CDM (black) models without Starobinsky's term. 
   Left panel corresponds to estimation without CMB data, right panel relates to estimation including CMB data.}
   \label{hubble}
\end{figure}

Finally, at the fig \ref{hubble} Hubble's diagrams for all our models  are shown and compared with $\Lambda$CDM.
Most of them are practically indistinguishable from the concordance model on the level of Hubble diagrams.
Moreover, the $\chi^2-$test gives mostly comparable results and in few cases priority belongs to new models provided
that CMB data are neglected. It is clearly seen that only for the special case $II_{\alpha=0}$
(fitting made with and without CMB data), the corresponding plots considerable differ from all remaining ones.
These motivates us for further analysis and employing Bayesian model selection methods.

\section{Models comparison and selection}

Bayesian theory enables to compare investigated models, i.e. enables to show which one is the best (most probable) in the light of analysed data. Let us consider the posterior probability for model indexed by $i$ ($M_i$):
\begin{equation}
 P(M_i|D)=\frac{P(D|M_i) P(M_i)}{P(D)}.
 \end{equation}
$P(M_i)$ is the prior probability for the model under investigation, $P(D)$ is normalization constant, $P(D|M_i)$ is the marginalized likelihood (also called evidence) and is given by:
\begin{equation}
P(D|M_i)= \int P(D|\bar{\Theta} , M_i) P(\bar{\Theta} |M_i) \mathrm{d}  \bar{\Theta}.
\end{equation}

It is convenient to choose the base model $M_b$ and compare all models from investigated set with respect to this base model by considering the ratio of posterior probabilities (posterior odds):
\begin{equation}
\frac{P(M_b|D)}{P(M_i|D)}=\frac{P(M_b)}{P(M_i)} \frac{P(D|M_b)}{P(D|M_i)}=\frac{P(M_b)}{P(M_i)} B_{bi}. \nonumber
\end{equation}
When all considered models are equally prior probable the posterior odds is reduced to the evidence ratio, so called Bayes Factor ($B_{bi}$). Its value can be simply interpreted as the strength of evidence in favour of base model : $0<\ln B_{bi} <1$ as inconclusive, $1<\ln B_{bi} <2.5$ as weak, $2.5<\ln B_{bi} <5$ as moderate and $\ln B_{bi} > 5$ as strong evidence. Since $\ln B_{bi} = -\ln B_{ib}$ the negative values of $\ln B_{bi}$ should be interpreted in favour of model under investigation.

In  Table~\ref{evidence} one can find values of logarithm of Bayesian Factor for new models, which were calculated with respect to the base $\Lambda$CDM model using CosmoNest code. The values were averaged from five runs.

As one can generally conclude models based on solution II are not supported by data used in analysis: we found weak and strong evidence in favor of $\Lambda$CDM model (for both with and without CMB data cases). When considering Model $I$ one can find out that there is weak evidence in
favor of it with respect to $\Lambda$CDM model in the light of observations from late Universe. When information from
early Universe is included, i.e. CMB data point is taken into account, the conclusion change: moderate evidence is found
in  favor of $\Lambda$CDM model. Amazingly, the special case of quadratic gravity ($I_{\beta =0}$) is not supported
by the data, in both cases the evidence in favor of the base model is strong. The result for model $I_{\alpha=0}$ is
inconclusive when the data from late Universe are considered. Admitting information from CMB gives us strong evidence
in favor of $\Lambda$CDM model.

\begin{table}[!ht]
\begin{center}
\begin{tabular}{| c | c | c | c | c |}
\hline
\hline
Model & \multicolumn{2}{|c||}{Estimation without CMB data} & \multicolumn{2}{|c||}{Estimation with CMB data} \\
\hline
 & $\ln B_{\Lambda CDM ,Model}$ & $\chi^2_{TOT}/2$ & $\ln B_{\Lambda CDM ,Model}$ & $\chi^2_{TOT}/2$\\
  \hline
$I_{\alpha=0}$ & $ -0.8 \pm 0.4 $ & $272.287$ & $453.3 \pm 0.6 $&  $706.235$\\
$I_{\beta=0}$ & $ 1665.2 \pm  0.3 $ & $274.142$ & $35.5 \pm 0.2 $ & $310.949$\\
$I$ & $ -2.2 \pm 0.3 $ & $271.400$ & $4.2 \pm 0.3 $ & $276.668$\\
$II_{\alpha=0}$ &$ 21.3 \pm 0.2 $ & $294.233$ & $169.1 \pm 0.3 $ & $438.839$\\
$II$ & $ 6.5 \pm 0.3 $ & $279.237$ & $206.6 \pm 0.3 $ & $475.881$\\
$\Lambda$CDM & 0 & $276.583$ & 0 & $276.726$\\
  \hline
\end{tabular}
\caption{Values of the logarithm of Bayesian Factor together with the corresponding $\chi^2/2$ for models based on solutions I and II, with respect to $\Lambda$CDM model.}
\label{evidence}
\end{center}
\end{table}

\section{Phase space descriptions of model dynamics}

Having fixed free parameters of our models on can think about their dynamical properties encoded in a time evolution
of the scale factor. In fact, the dynamics of our models is determined by very complicated  Friedmann type equations
 which constitute  first order ordinary (non-linear) autonomous differential equations on the scale factor.
 Moreover  r.h.s. of the generalized Friedmann equation (\ref{H_H0_2rozw},\ref{H_H0}) is rather rational
 than polynomial function (cf. (\ref{sfe})) of $a$. While obtaining exact solutions of these equations is very difficult
 (if possible at all) it is enough to apply qualitative and numerical methods of analysis of differential equations.
 The main goal of this method is instead of studying individual trajectories of the system under consideration to
 describe a geometrical structure of a phase space. The main advantage 
is the possibility of investigating entire evolution, represented by trajectories
in the phase space, for all admissible initial conditions. The phase space
(the phase plane $(a, \dot a)$ in our case) is organized through critical
points and trajectories. As a result we obtain a global phase portrait of
the system illustrating the stability of special solutions as well as
its generic  properties among all evolutional paths in the phase space.

It would be useful to represent the dynamics of the cosmological models in
terms of dynamical system theory. For this aim let us re-parametrize the original time
variable in (\ref{H_H0_2rozw},\ref{H_H0}) to the new re-scaled variable, say $\tau$ such that $d\tau=|H_0|dt$ and define
an effective potential function $V(a)=-\frac{1}{2}a^2K(a)G(a)$. Then we obtain that dynamics is reduced to the dynamics
of a fictitious particle of unit mass moving on a half line in the effective potential $V(a)$ with energy level
$E_\kappa=-\kappa/2$, where $\kappa= 0, \pm 1$ is the spatial curvature index (for generality one considers models which
are not flat):
\begin{equation}
\frac{1}{2}\left( \frac{da}{d \tau} \right)^2 + V(a) = E_\kappa
\end{equation}
The system is defined in the part of the configuration space in which
$E_\kappa-V$ is non-negative. 
Therefore the dynamics of cosmological model is governed by  
dynamical system
\begin{align}
\frac{da}{d\tau} &= x \\
\frac{dx}{d\tau} &= - \frac{dV}{da}.
\end{align}
of Newtonian type. Critical points of this  system correspond
to extremes on a diagram of the effective potential. Because the stability of the
critical point depends on the eigenvalues of a linearization matrix (solutions
of the characteristic equation) calculated at this point one obtains two
possibilities (a trace of the linearization matrix is always vanishing). If one has a maximum then the critical point is a saddle. In contrast  a minimum on the diagram of the potential function means that we
have a center type of critical point. Critical points provide stationary solutions: stable for
minimum and unstable otherwise.

In fact dynamics of the system can be read off from the graph of potential function itself (in cosmological setting the
effective potential is non-positive: $V\leq 0$). Since conservation of (total)
energy holds, each trajectory is labeled by some energy level $E=constant$ (might be negative). The difference $E-V\geq 0$ is equal to kinetic energy and provides information about the velocity of cosmic expansion: increasing values of the
potential are slowing down the evolution. Because of this the region
below potential plot is forbidden for the motion.  Particularly,  points on the graph are turning points. For example,
models with a classical bounce contract to some minimal value of the scale factor and then after approaching this
value expand (to infinity). Oscillating models are, therefore, result of a potential cavity. As a result one can
simply obtain different types of evolutions admissible by different values of energy levels.
For example,  as it was mentioned above, different choice of spatial curvature index $\kappa$
will determine different cosmic evolution. Such visualization is complementary to the one which offers
phase portrait.

\FloatBarrier
\begin{figure}[h!!!!]
\centering
   \includegraphics[width=0.496\textwidth]{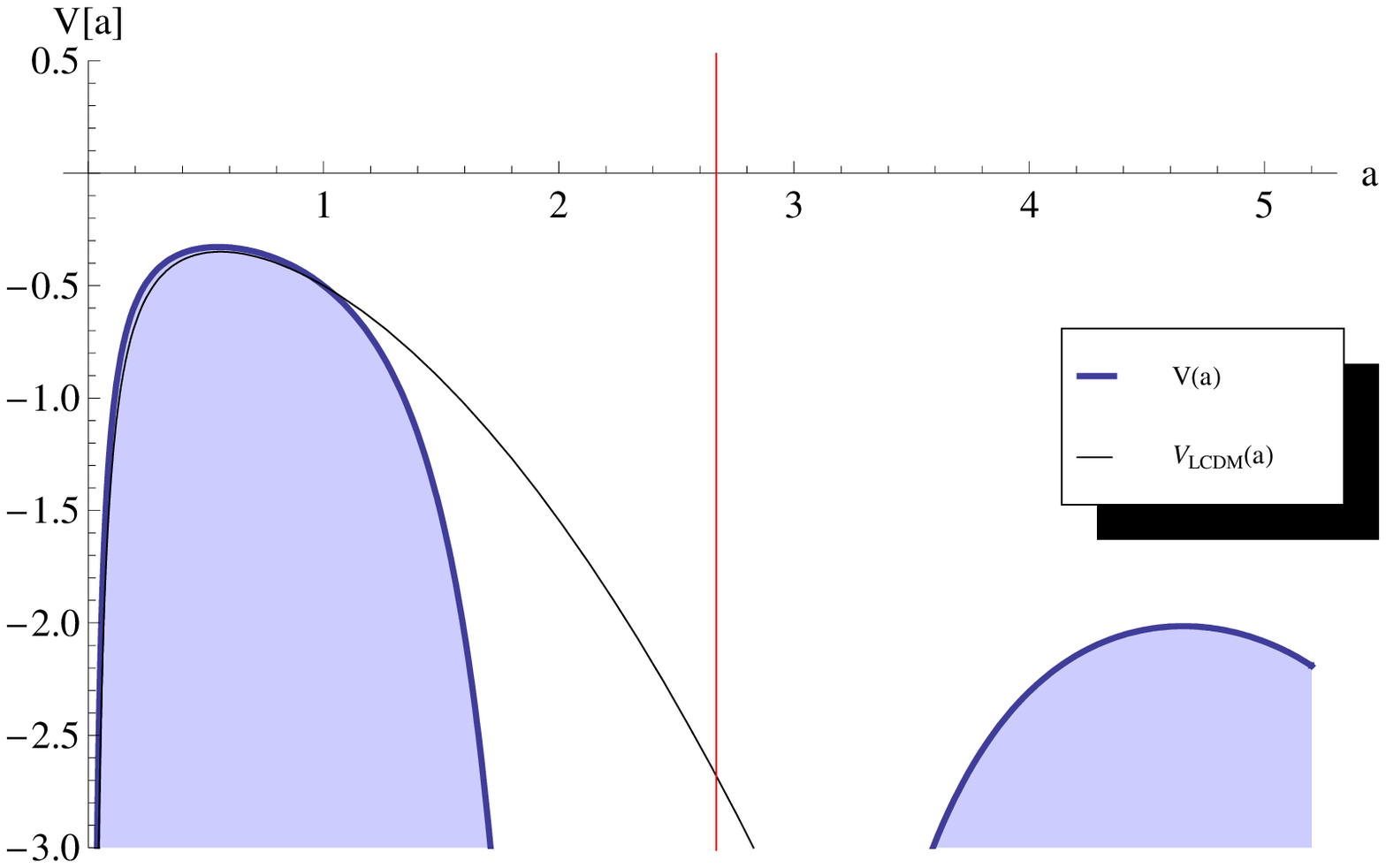}
   \includegraphics[width=0.496\textwidth]{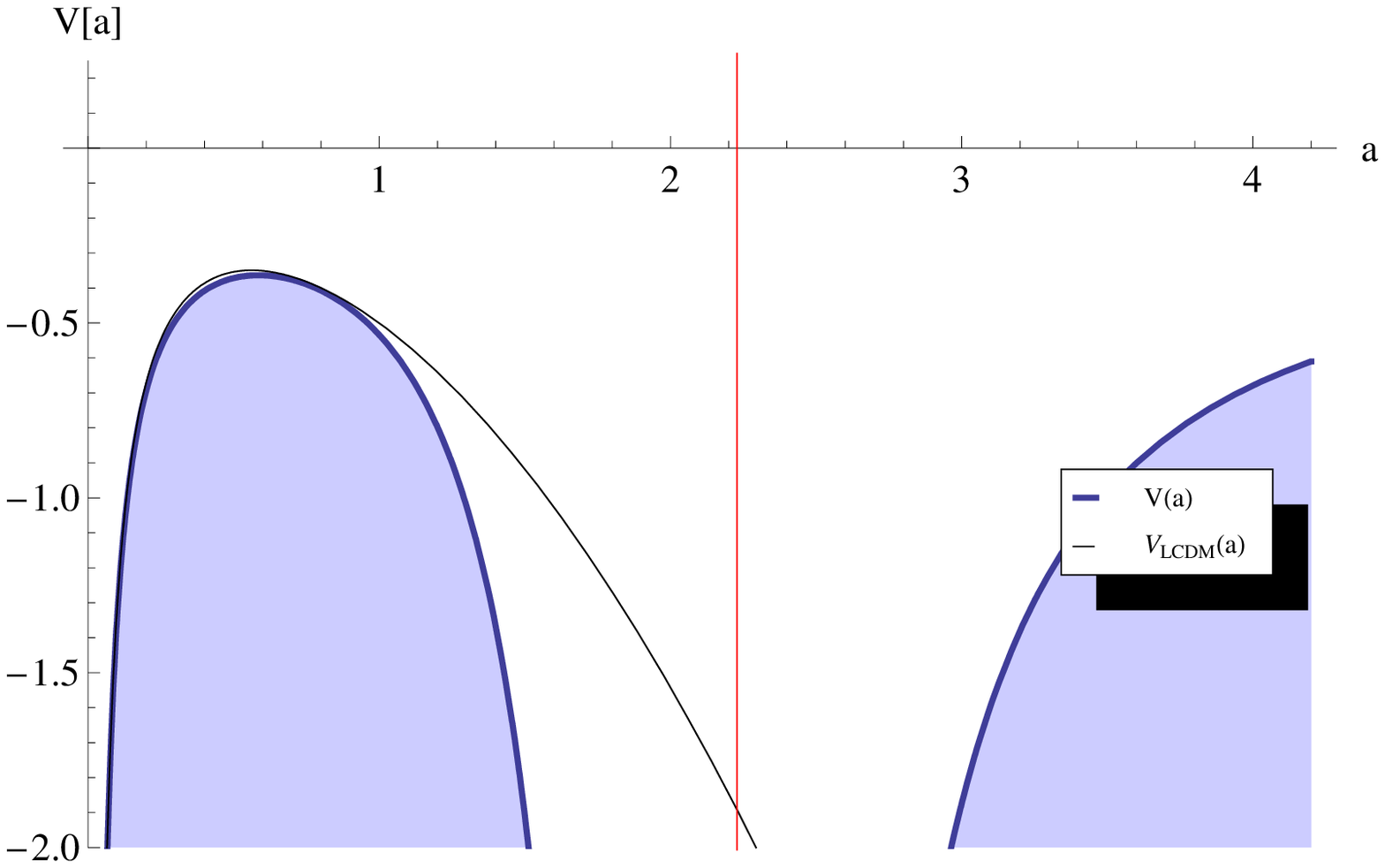}\\
   \caption{The diagram of the effective potential in particle--like representation of cosmic dynamics
   for model I versus $\Lambda$CDM model (left picture relates to estimation without
   CMB data, the right relates to estimation employing CMB data; table \ref{zestawienie}, No 3,  6, 9, 12).
    Note that till the present epoch two potential plots almost coincide. Particulary, one can observe decelerating BB era.
    Maximum of the potential function corresponds to Einstein's unstable static solution (saddle point). Discrepancies become important
    in the future time: e.g. discontinuities of the potential functions (vertical, red lines) denote that $V \rightarrow -\infty$, i.e.
   $\dot a \rightarrow \infty$ for $a \rightarrow a^{final}$. It turns out to be finite--time (sudden) singularity.
   In any case the shadowed region below the graph is forbidden for the motion.}
   \label{potencjal_ii}
\end{figure}

\begin{figure}[h!!!!]
\centering
   \includegraphics[width=0.496\textwidth]{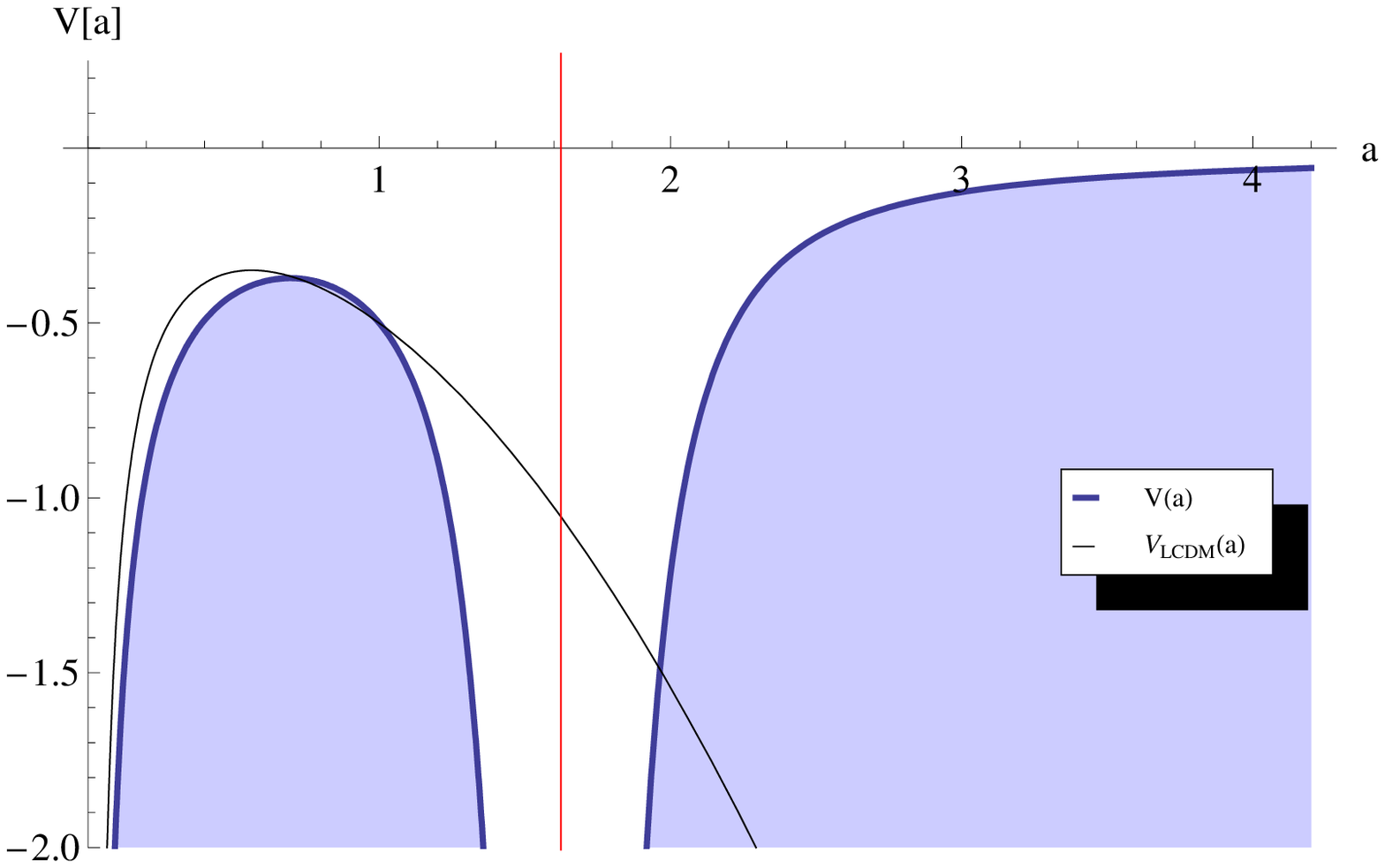}
   \includegraphics[width=0.496\textwidth]{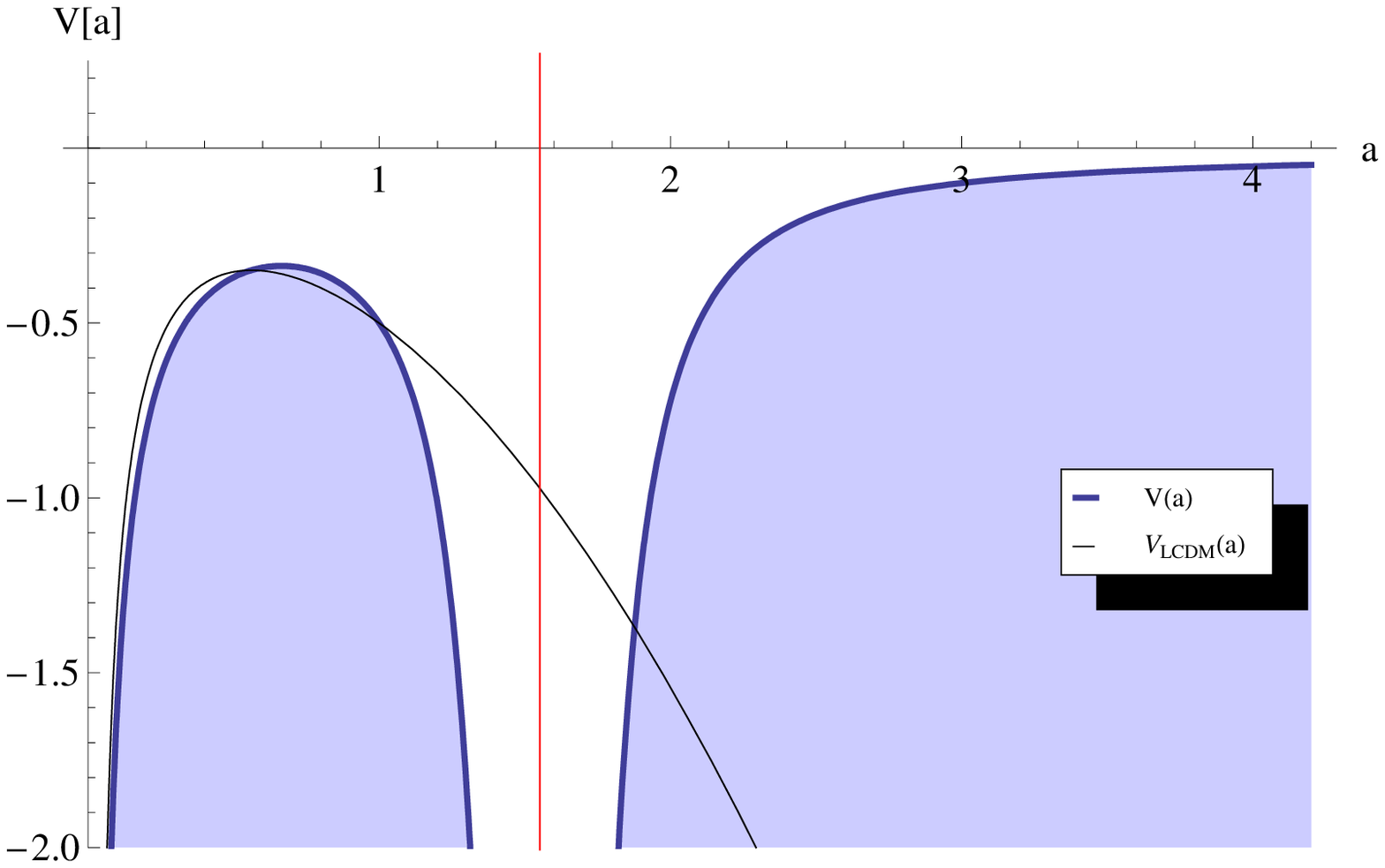}\\
   \caption{The diagram of the effective potential in particle like representation of cosmic dynamic
   for the model of quadratic gravity $I_{\beta=0}$ versus $\Lambda$CDM model
   (left picture relates to estimations without CMB data, the right relates to estimations employing CMB data;
   Table \ref{zestawienie}, no 2, 6, 8, 12).  Maximum of the potential function corresponds to unstable static solution
   (saddle point). Again, until the present epoch there is no striking differences between plots.
    One can observe finite--size sudden singularity in the near future (vertical, red lines).
    In any case the shadowed region below the potential is forbidden for the motion.}
   \label{potencjal_ob_0_ii}
\end{figure}

\FloatBarrier

\begin{figure}[h!!!!]
\centering
   \includegraphics[width=0.496\textwidth]{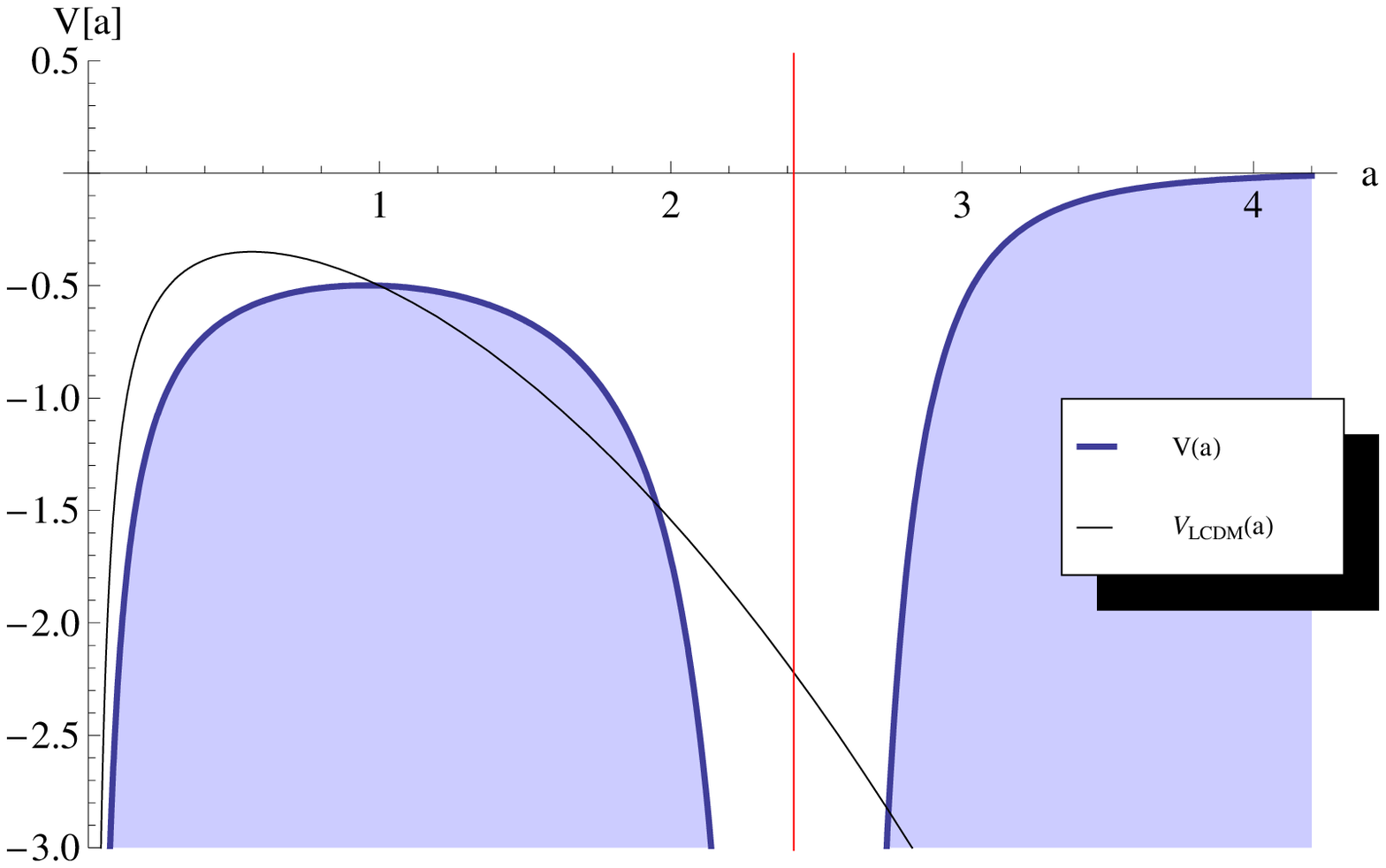}
   \includegraphics[width=0.496\textwidth]{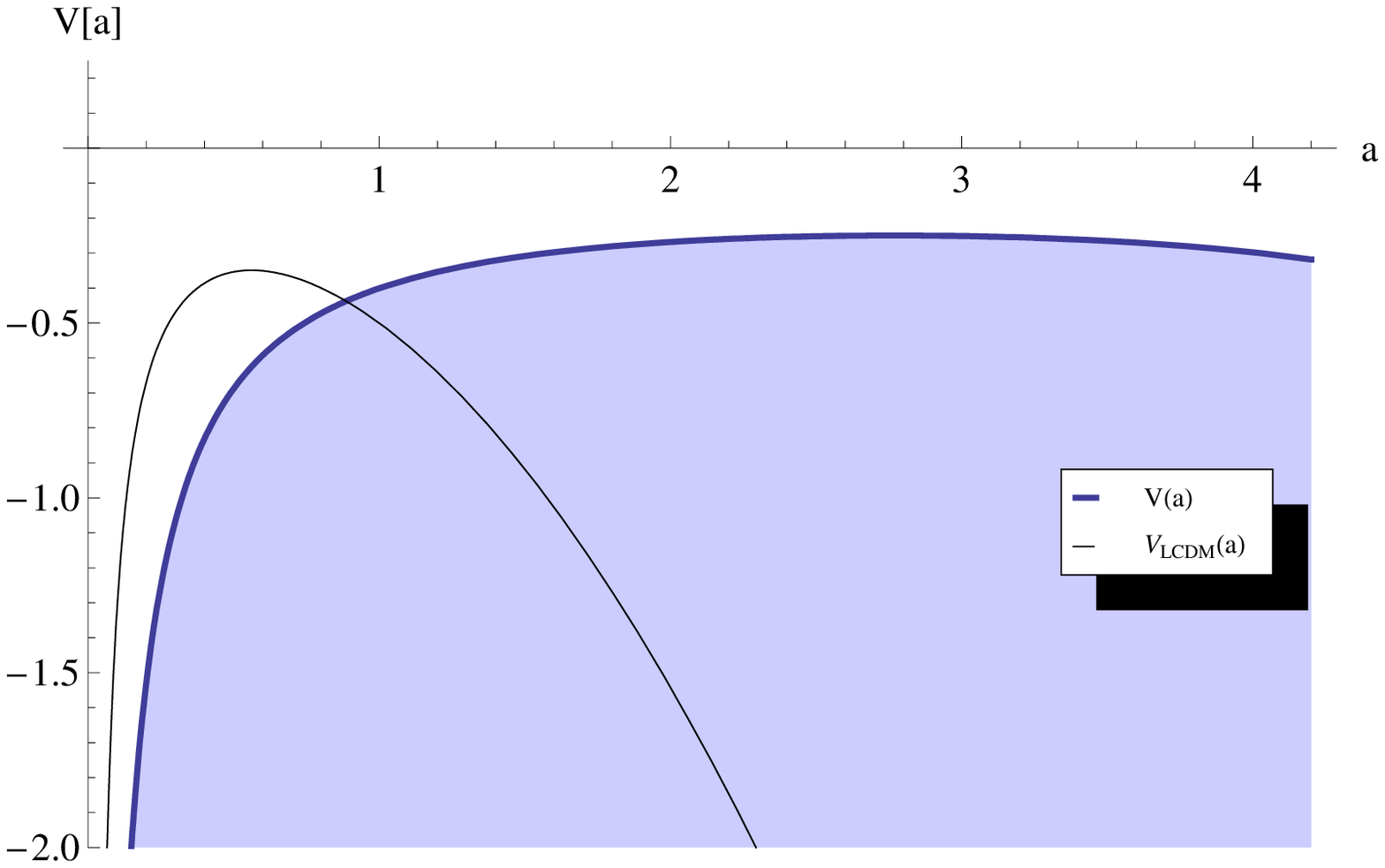}\\
   \caption{The diagram of the effective potential in particle like representation of cosmic dynamic
   for the model $I_{\alpha=0}$ versus $\Lambda$CDM model
   (left picture relates to estimations without CMB data, the right relates to estimations employing CMB data;
   Table \ref{zestawienie}, no 1, 6, 7, 12).  Both pictures represent decelerating, till the present epoch,
   universe with BB initial singularity.   Now discrepancies with $\Lambda$CDM model are more evident.  In any case the shadowed region below the
   potential is forbidden for the  motion.}
   \label{potencjal_oa_0_ii}
\end{figure}

\begin{figure}[h!!!!]
\centering
   \includegraphics[width=0.496\textwidth]{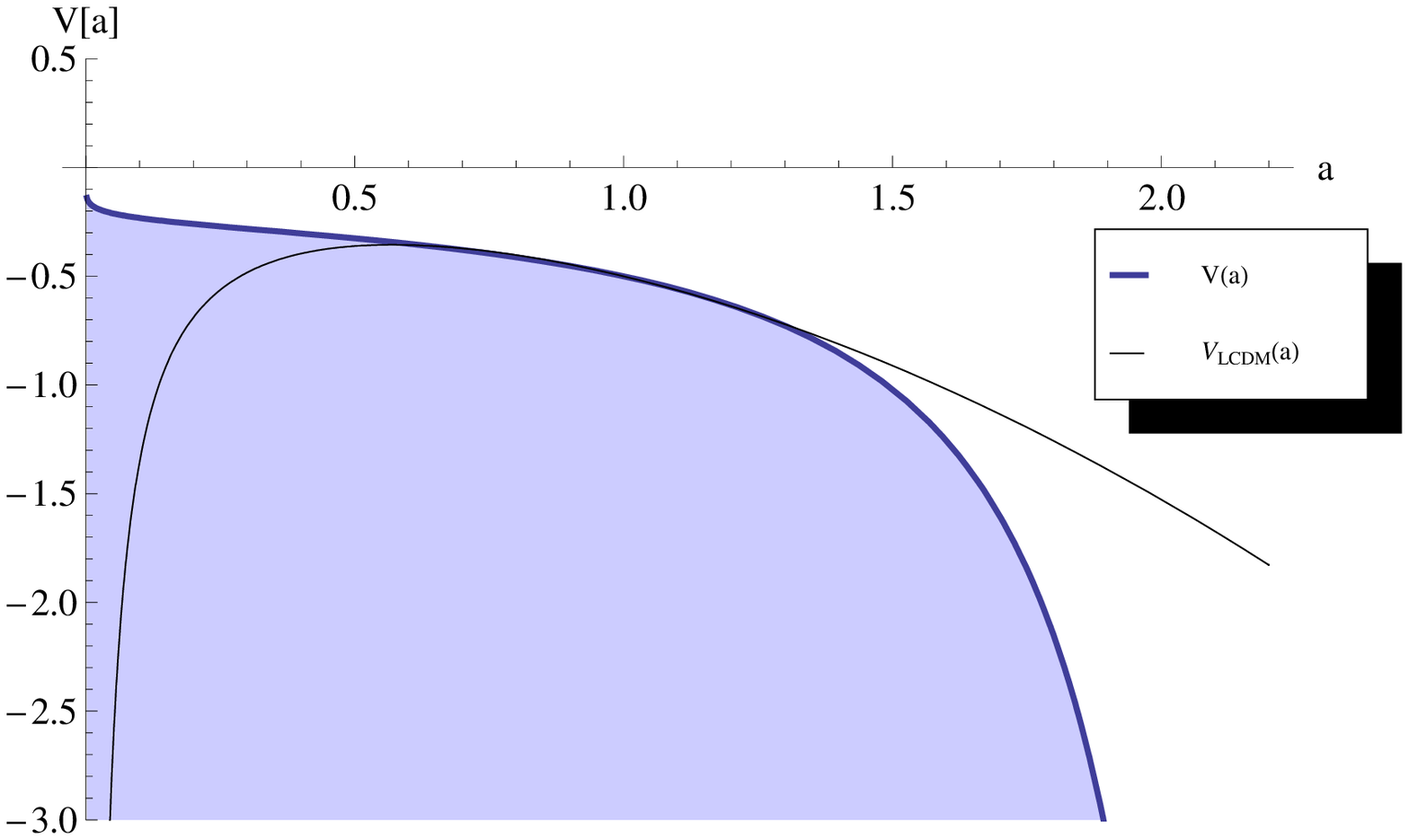}
   \includegraphics[width=0.496\textwidth]{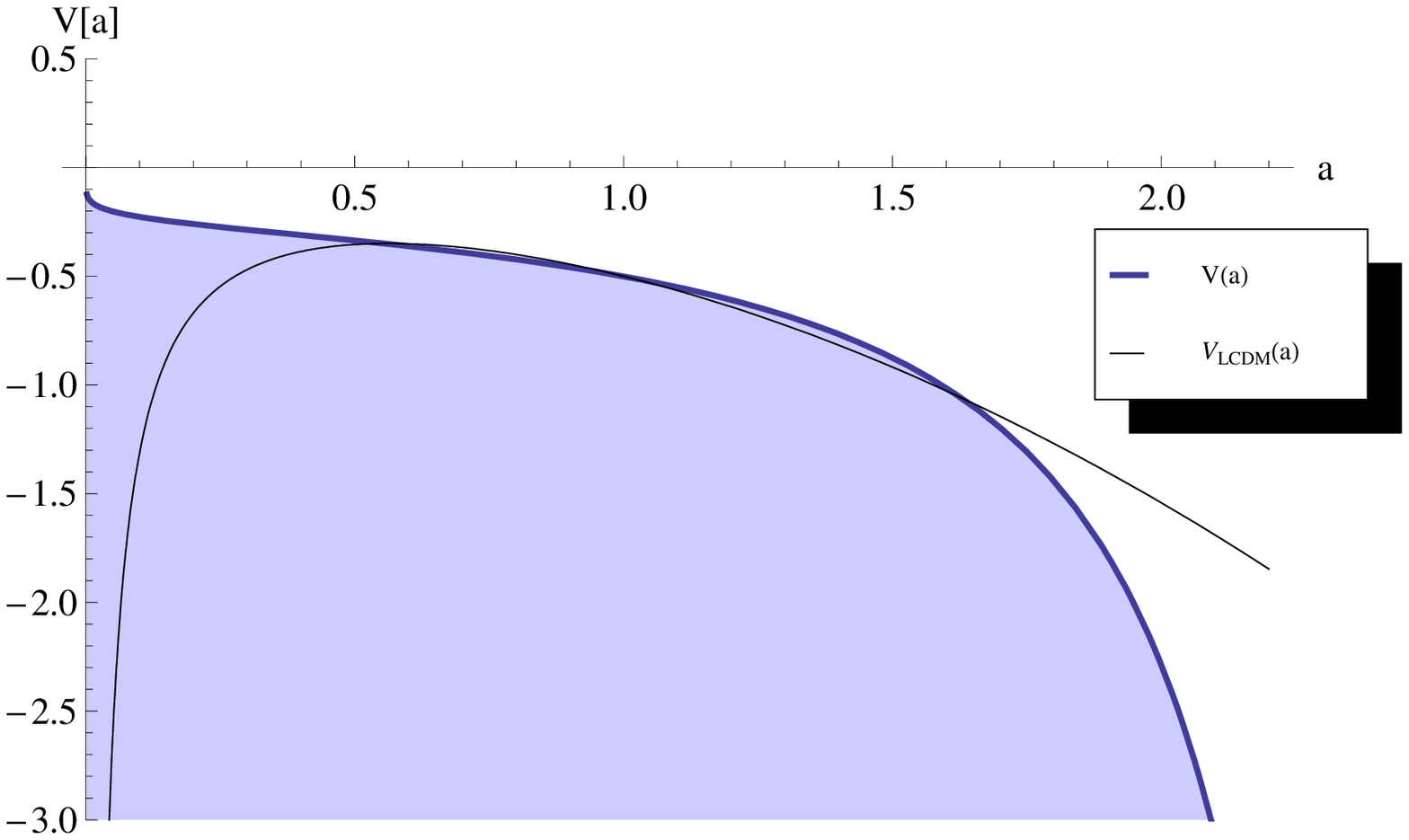}\\
   \caption{
   The diagram of the effective potential in particle--like representation of cosmic dynamics for model II
   versus $\Lambda$CDM model (left picture relates to estimation without  CMB data, the right
   relates to estimation with CMB data; Table \ref{zestawienie}, no 5, 6, 11, 12). The evolution of
   the model is represented through the energy level. Therefore both models are bouncing type. The universe is
   contracting, reaches the minimal size and then is expanding with acceleration. Notice that some part of the
   potential plot coincide with $\Lambda$CDM model.}
   \label{potencjal_solution_i}
\end{figure}
\FloatBarrier

\begin{figure}[h!!!!]
\centering
   \includegraphics[width=0.496\textwidth]{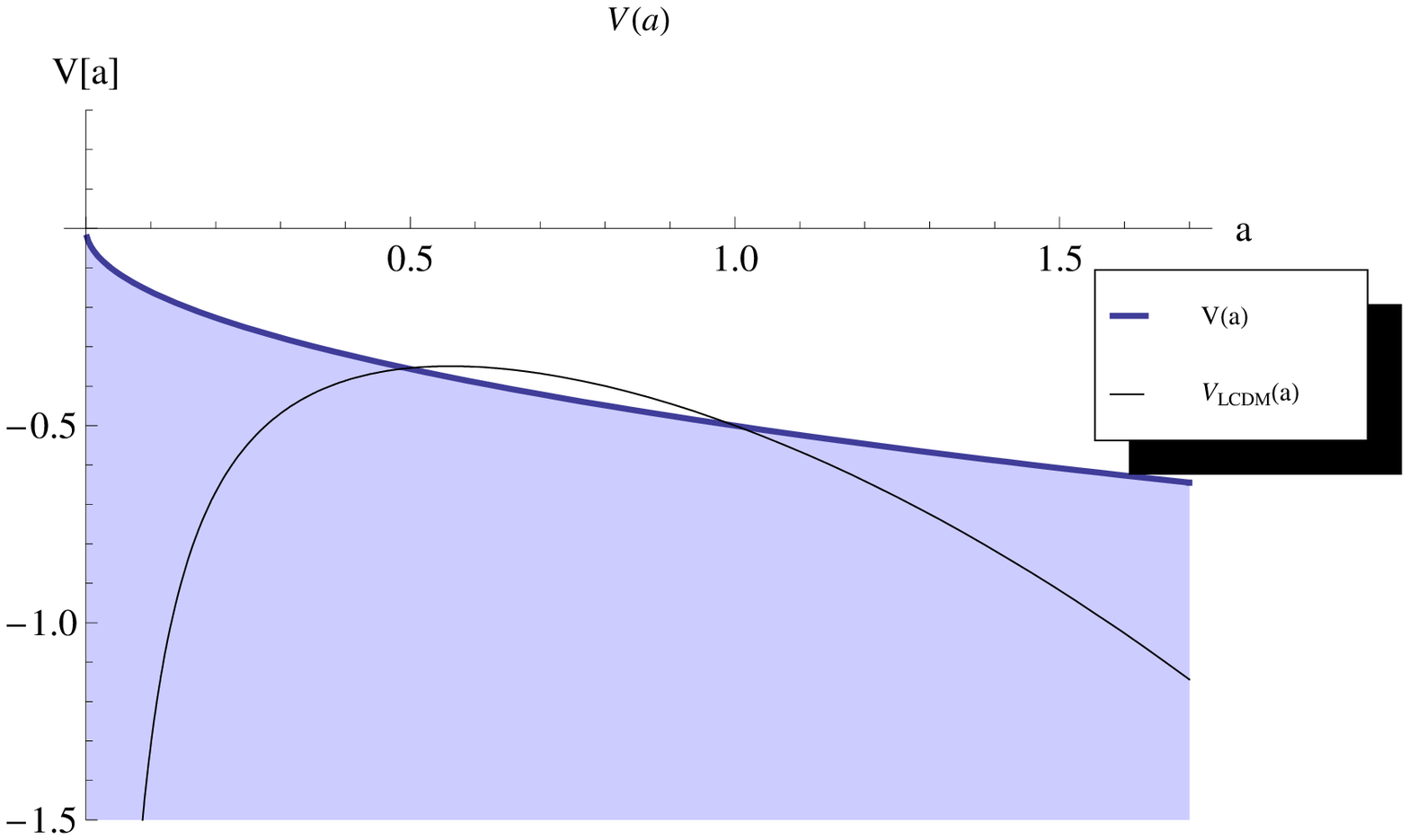}
   \includegraphics[width=0.496\textwidth]{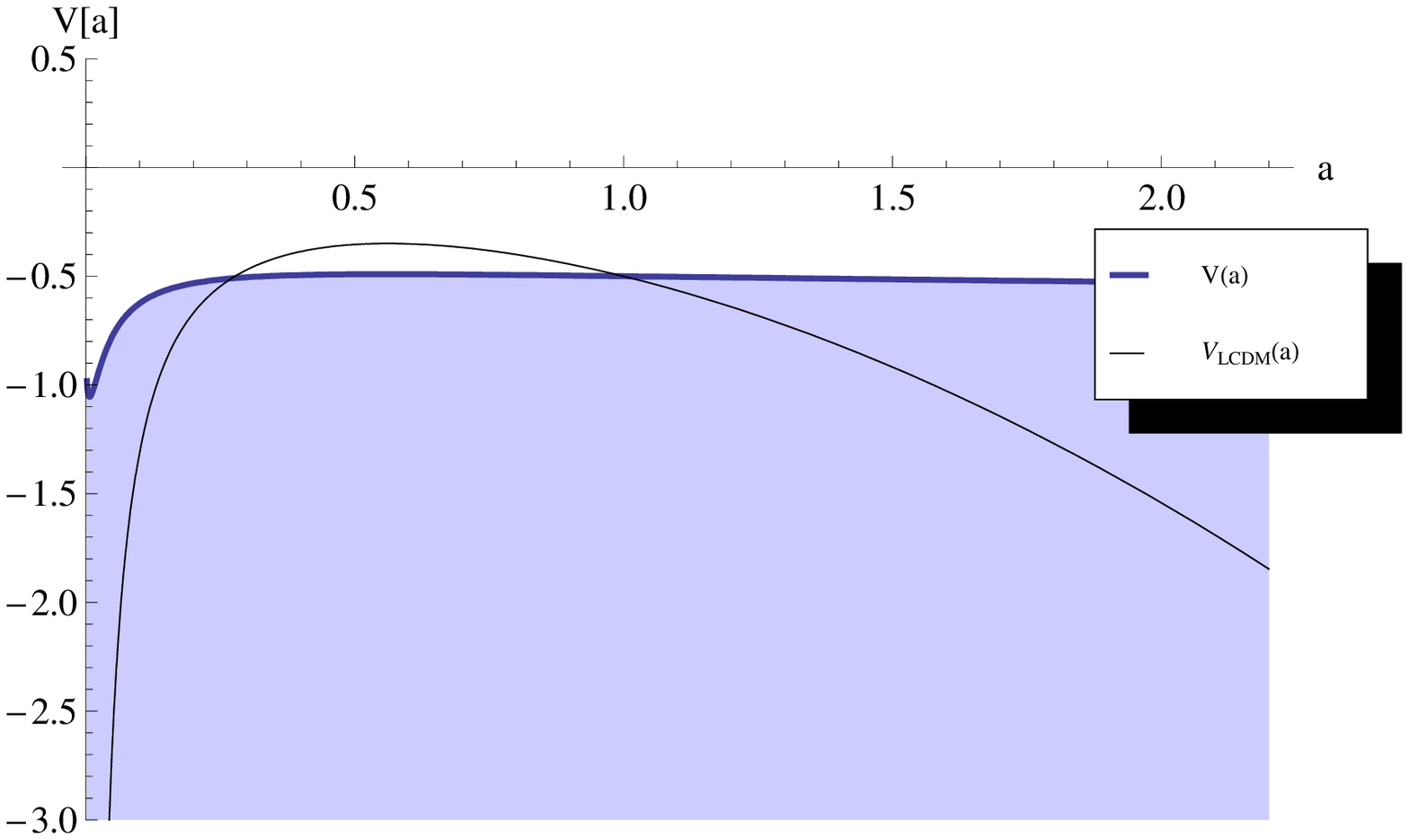}\\
   \caption{The diagram of the effective potential in particle like representation of cosmic dynamics
   for the model $II_{\alpha=0}$ versus $\Lambda$CDM model (left picture relates to estimation without
   CMB data, the right relates to estimation with CMB data; Table \ref{zestawienie}, no 4, 6, 10, 12). One
   can observe big discrepancies with respect to the concordance model. Moreover, for the first time, adding CMB data changes the type of expansion from accelerating to not accelerating.}
   \label{potencjal_oa_0}
\end{figure}
\FloatBarrier

\begin{figure}[h!!!!]
\centering
   \includegraphics[width=0.60\textwidth]{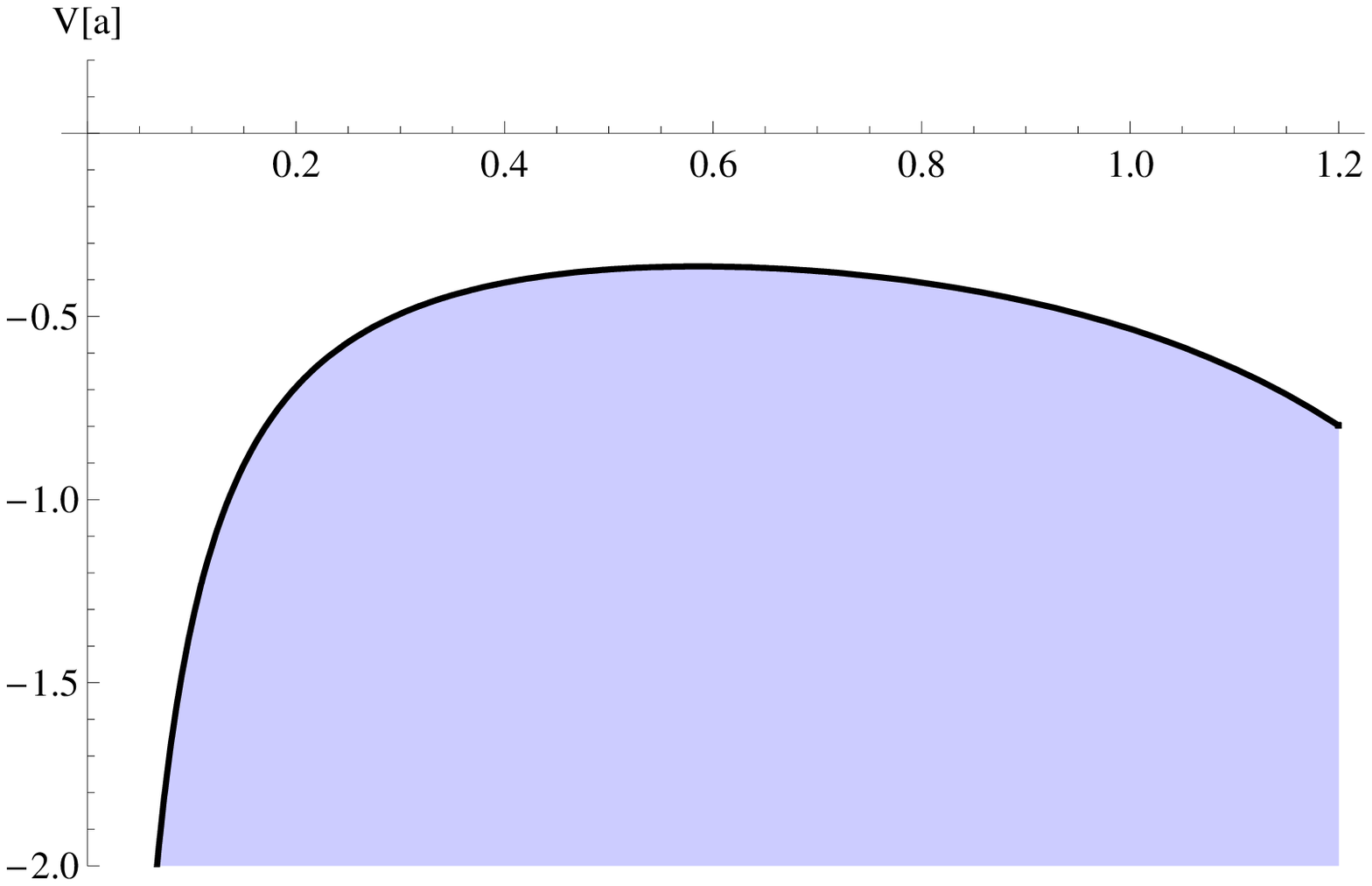}\\
   \includegraphics[width=0.55\textwidth]{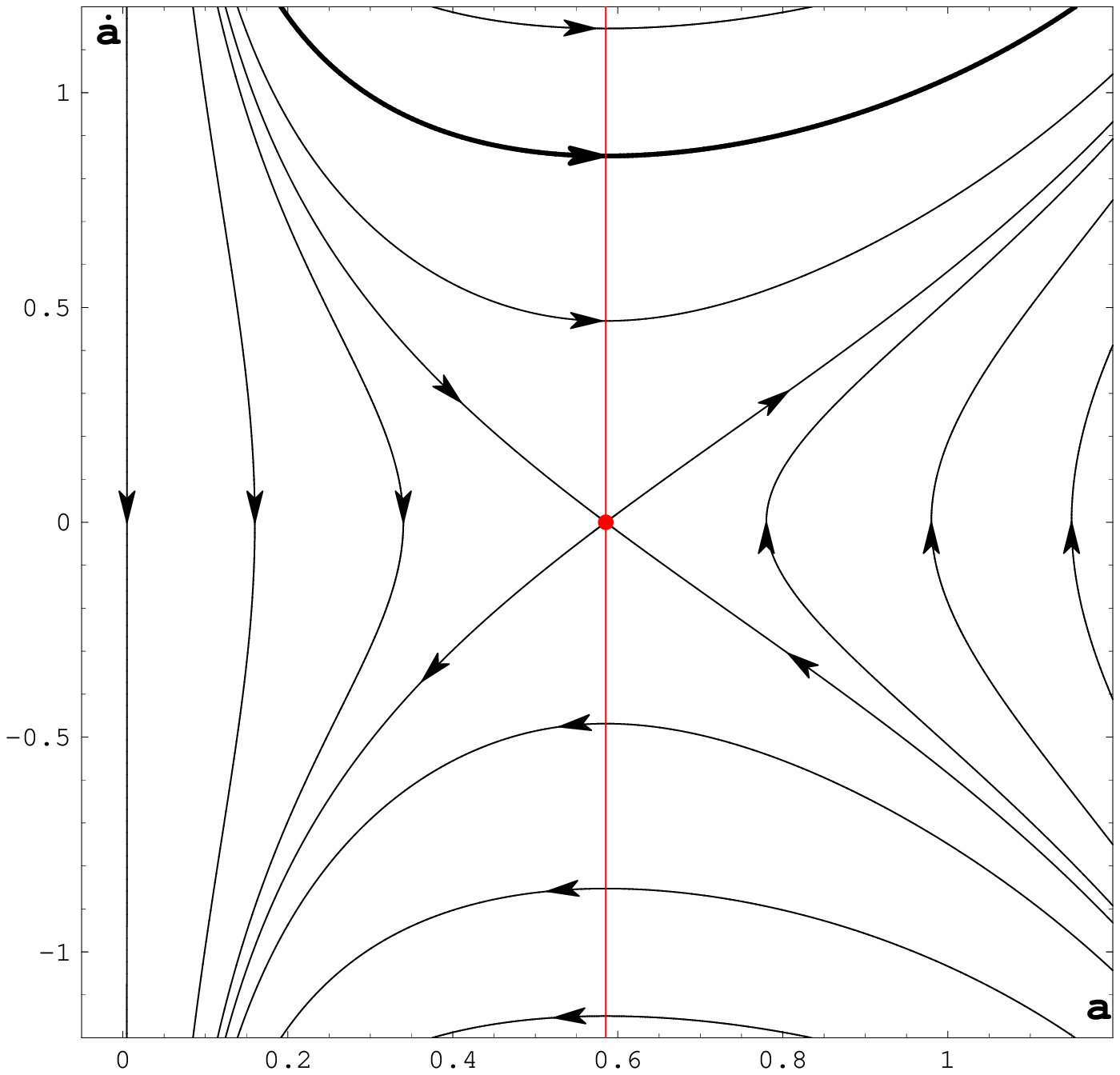}\\
   \caption{The diagram of the effective potential and the corresponding phase portrait of the model $I$
   for estimations using CMB data. However for estimations without CMB data the phase portrait looks
   similar.
   We marked as a bolded trajectory of the flat model determined by the energy constraint $E=0$.
   The phase space is divided by this trajectory  on two domains at which lie closed and open models.
   We have situated the only possible critical point, the saddle
   $(a_{\text{static}},0)$ which represents the Einstein static universe. The vertical red line $a=a_{\text{static}}$
   passing through the saddle critical point divides  each trajectory into two parts:
   decelerating ($V(a)$ is a growing function of its argument) and accelerating ($V(a)$ is
   a decreasing function of the scale factor) eras. This part of the phase portrait is topologically  equivalent to
   phase portrait of $\Lambda$CDM model. Particulary, the Bing Bang era is decelerating.}\label{cmb_portret_fazowy_ii}.
\end{figure}

\begin{figure}[h!!!!]
\centering
   \includegraphics[width=0.55\textwidth]{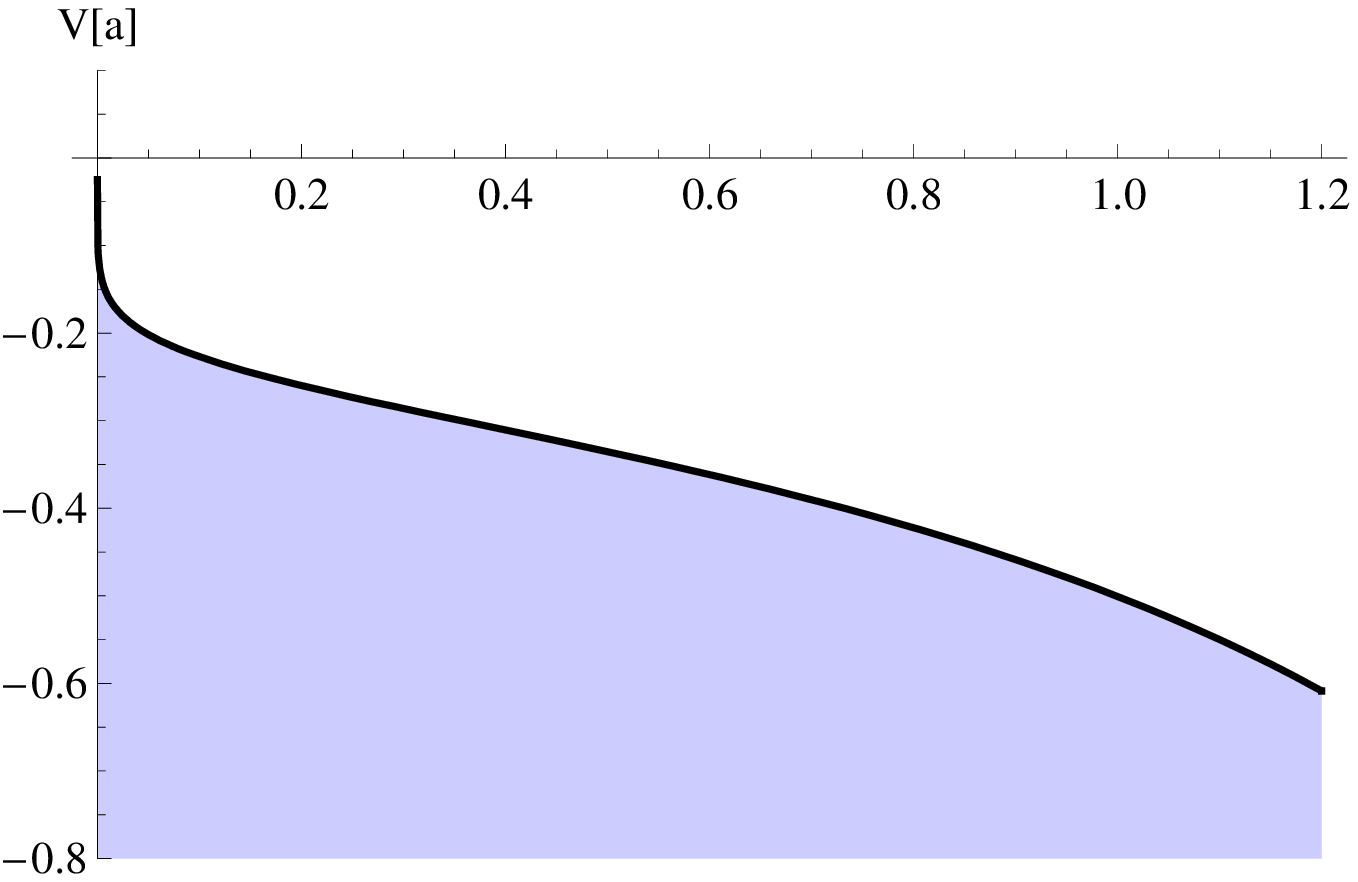}\\
   \includegraphics[width=0.55\textwidth]{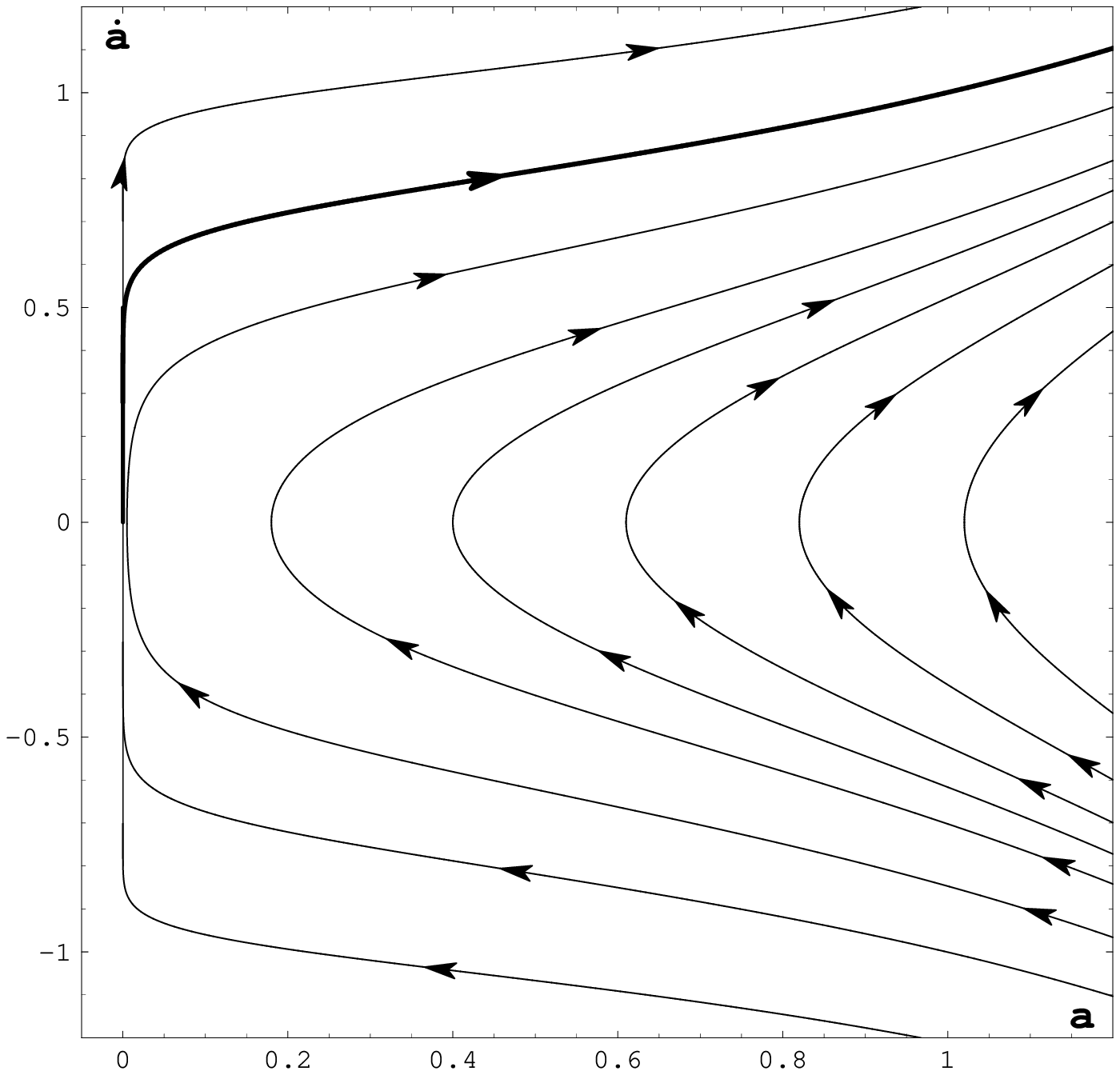}\\
   \caption{The diagram of effective potential and the corresponding phase portrait on the plain $(a, \dot a)$ for
   model $II$ -- estimations including CMB data. (In the case without CMB data the phase portrait looks similar.)
   The phase portrait shows all evolutional paths for all admissible
   initial conditions. We mark as bolded the trajectory of the flat model determined by the energy constraint $E=0$.
   The phase space is divided by this trajectory on two domains at which lie closed and open models.
   Note that all solutions are bouncing type.
   In contrast to $\Lambda$CDM model on can notice accelerating scenario from the very beginning.}\label{cmb_portret_fazowy_i}.
\end{figure}

On the diagrams  we have illustrated the effective potential functions (Fig. \ref{potencjal_ii}, \ref{potencjal_oa_0_ii},
\ref{potencjal_ob_0_ii}, \ref{potencjal_solution_i}, \ref{potencjal_oa_0}) as well as generic
samples of phase portraits for our models (Fig. 
\ref{cmb_portret_fazowy_ii}, \ref{cmb_portret_fazowy_i}). We found that although analytical formulae for
effective potential functions for various model are very different their numerical values and therefore a shape of
the corresponding graph might be almost the same. As it was explained above similar shapes give rise to similar
dynamical behavior. This is what happens in our case.

\FloatBarrier

\FloatBarrier

Let us discuss the plots of the effective potentials for new models in more details.  The model $I$ has reached
very good numerical agreement with $\Lambda$CDM for a wide range of the scale factor $a$: from the very beginning
till the present time (fig. \ref{potencjal_ii}). After Big Bang the universe expansion slows down (deceleration epoch)
till some critical value (equal to static solution) and since then speeds up. This cosmic acceleration lasts forever for the case of $\Lambda$CDM. In contrast, the model $I$ predicts finite size future singularity since the effective potential function (as well as velocity plot) has a pole at $a^{final}_I=2.359$. It can be shown that this pole has to be reached at finite time. Therefore it is known as a sudden future singularity \cite{sudden}. In the case
of quadratic gravity  ($I_{\beta=0}$), the shape of the potential function is the same as for model $I$ with poorer
numerical coincidence with $\Lambda$CDM (fig. \ref{potencjal_ob_0_ii}): $a^{final}_{I_{\beta=0}}=1.639$.
The initial deceleration era characteristic for $\Lambda$CDM model is absent only for the case of model $II$ which
offers a model of permanently accelerating universe of bouncing type. After a short period of strong acceleration
(possible inflation) there is an era of low acceleration followed by $\Lambda$CDM epoch. A characteristic for models
$I$ and $\Lambda$CDM saddle point is not present for the case of model $II$. Finite size future singularity of the potential function appears at
$a^{final}_{II}=2.343$. However on the Hubble diagrams differences between $\Lambda$CDM  and  models $I$, $I_{\beta=0}$ and $II$,  are negligible.
Similarity to $\Lambda$CDM is lost for both models with $\alpha =0$ (see fig. \ref{potencjal_oa_0_ii},
\ref{potencjal_oa_0}) which makes this case difficult to accept. 
Models of modified gravity enlarge zoo of possible cosmic evolutions because of the presence of
finite time final size singularity of the potential function at some
finite value of the scale factor.
 
\section{Other analysis and diagnostics}

\begin{figure}[h!!!!]
\centering
   \includegraphics[width=0.496\textwidth]{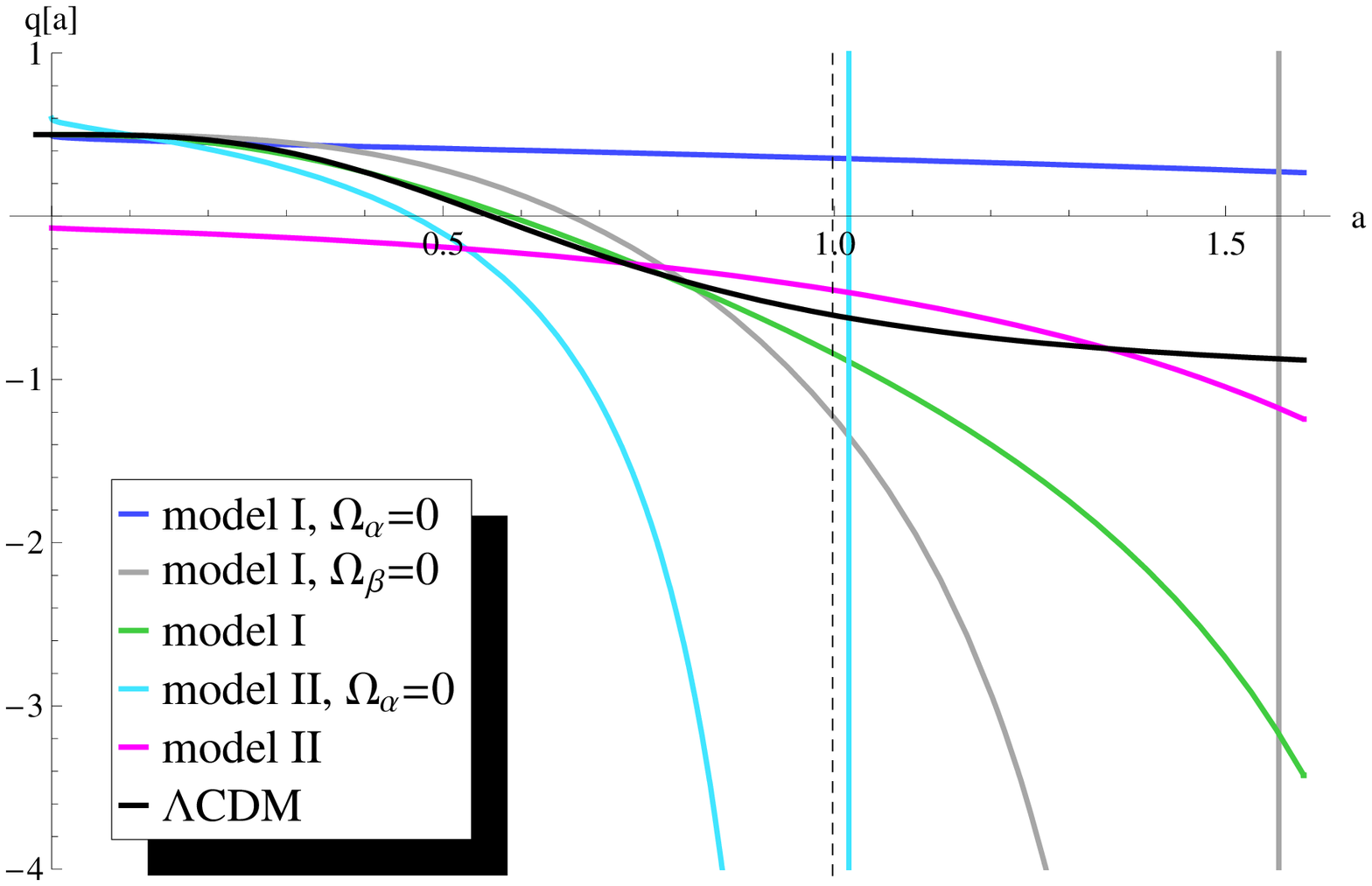}
   \includegraphics[width=0.496\textwidth]{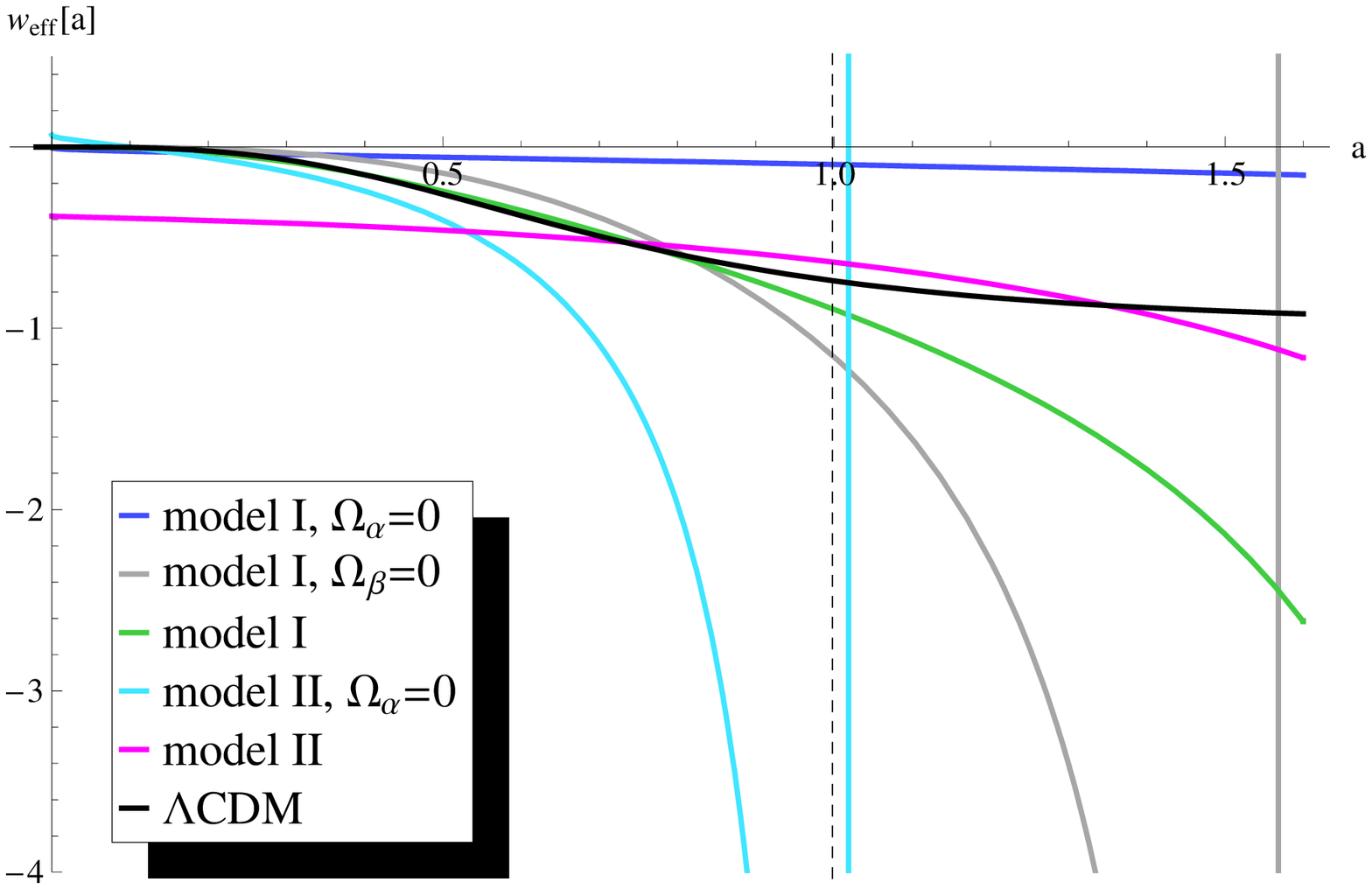}\\
   \caption{Plots of the deceleration parameter $q(a)$ (left panel) and the effective equation
   of state $w_{eff}(a)$ (right panel) for all models under investigation.
   Only model $II$ (magenta) provides permanent acceleration.
   Models $I_{\alpha=0}$ (blue) and $II_{\alpha=0}$ (light blue)
   have no acceleration epoch at all. There is intriguing intersection near $a= 0.75$
   for plots representing four models:
   $I$, $I_{\beta=0}$, $II$ and $\Lambda$CDM. Thin, vertical line denotes present time.
   The model parameters were fitted using SN, $H_z$, BAO and CMB data (Table \ref{zestawienie}, No 7--12).}
   \label{cmb_deceleration}
\end{figure}

\begin{figure}[h!!!!]
\centering
   \includegraphics[width=0.496\textwidth]{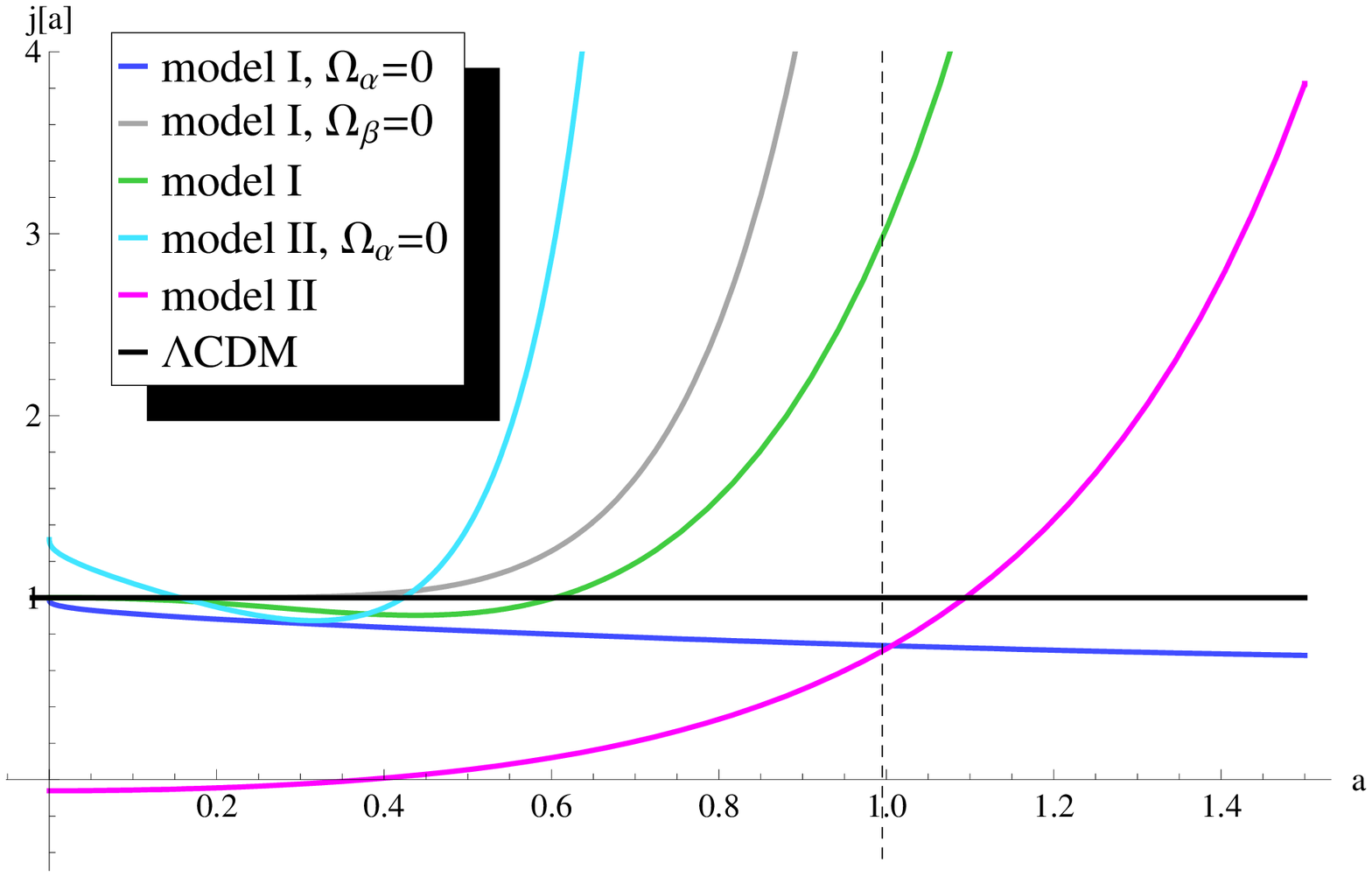}
   \includegraphics[width=0.496\textwidth]{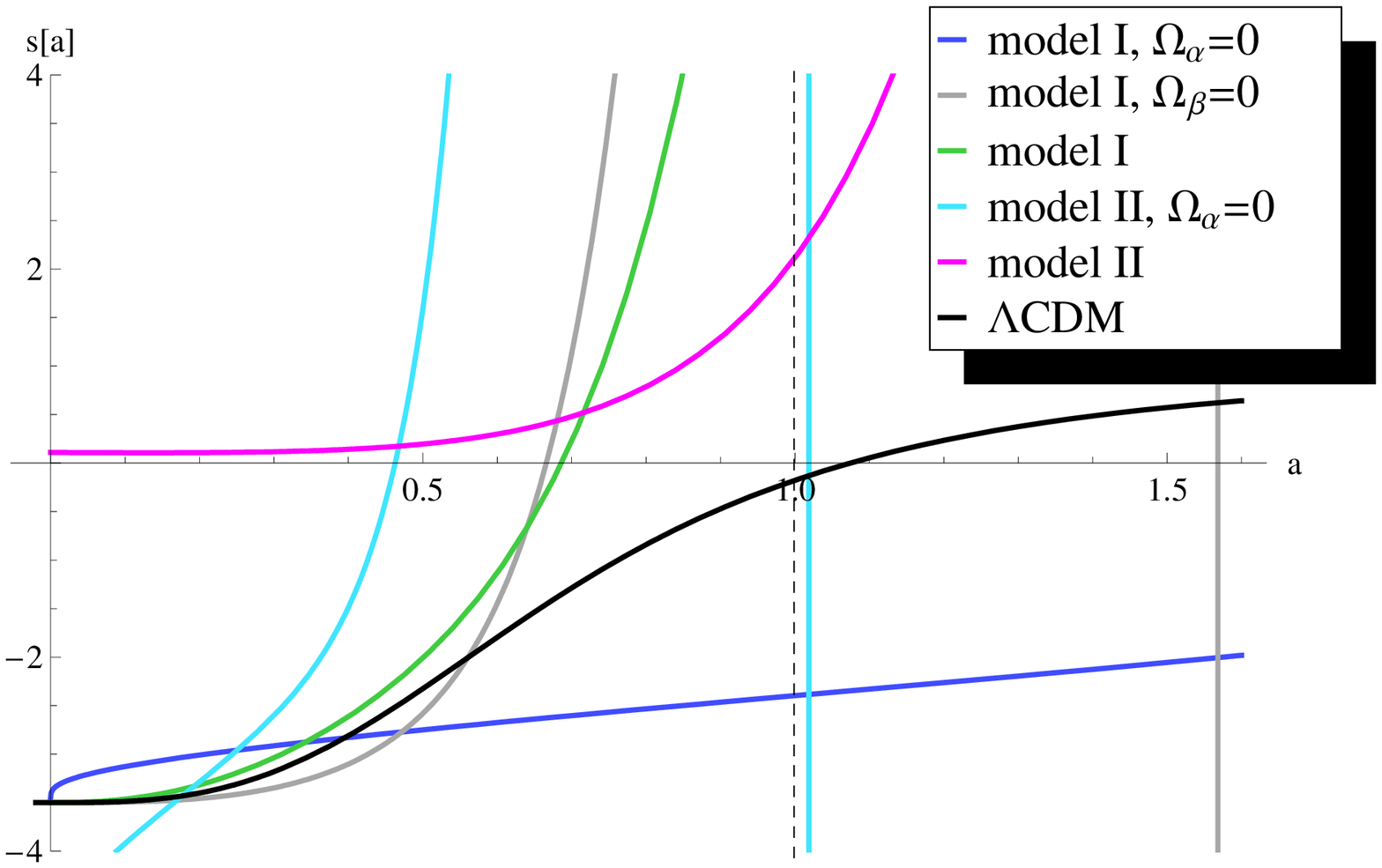}\\
   \caption{Plots of parameters JERK $j(a)$ (left panel) and SNAP $s(a)$ (right panel) for
   investigated models.
   From the figure one can see that
   different models predict different present-day values of JERK and SNAP. Unfortunately
   estimations of these parameters are beyond our present observational possibilities.
   However some recent analysis support the current values of jerk bigger than 2 \cite{jerk}.
   Thin, vertical line denotes present epoch. The model parameters were fitted using
   SNIa, $H_z$, BAO and CMB data (Table \ref{zestawienie}, No 7--12).}
   \label{snap_jerk}
\end{figure}

\begin{figure}[h!!!!]
\centering
   \includegraphics[width=0.496\textwidth]{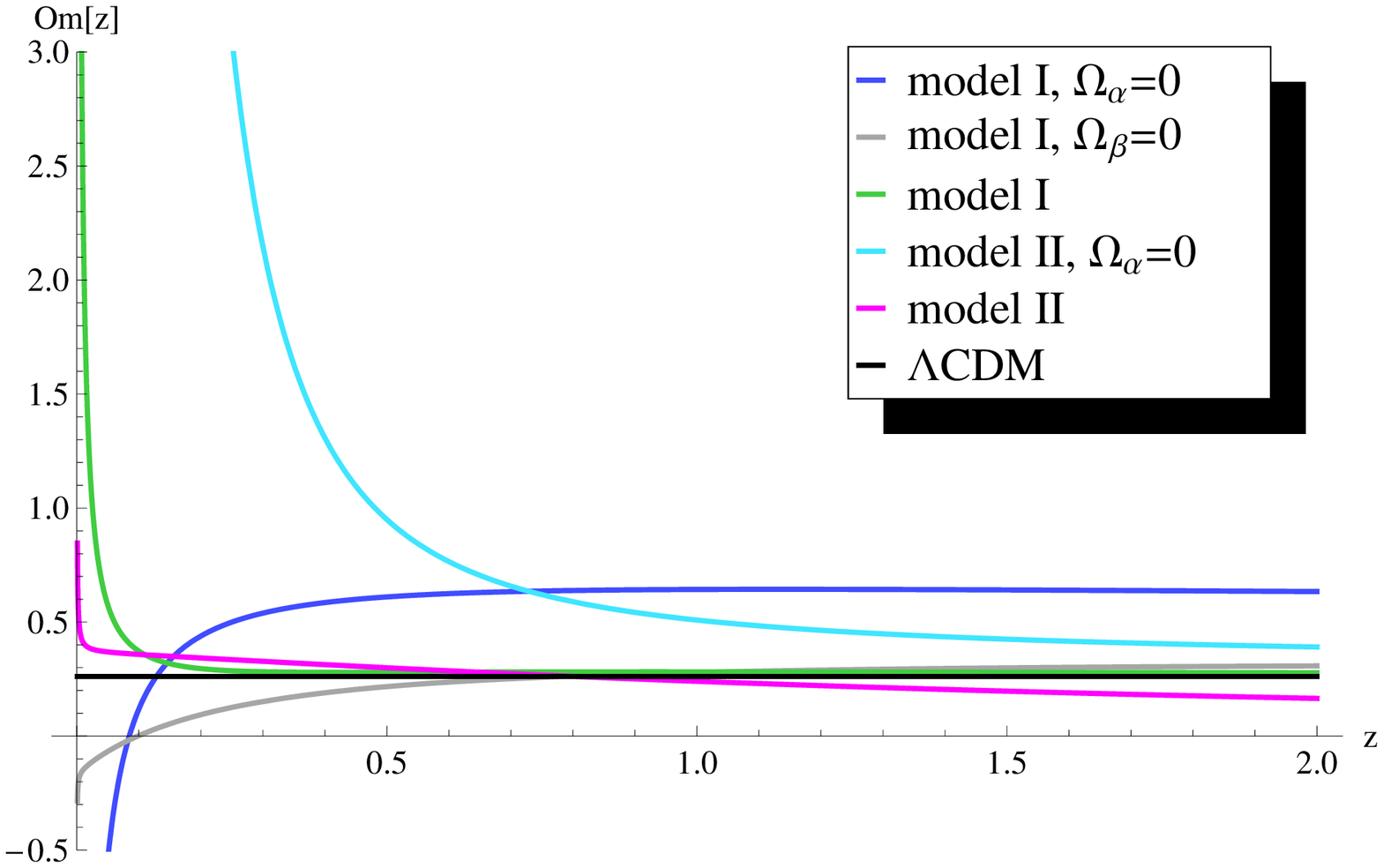}\\
   \caption{Plots of parameter $Om(z)$ allowing to answer the question on nature of dark energy \cite{Sahni}.
    For $\Lambda$CDM (black line) we have $Om(z)=constant$  which means that cosmological constant is the best
    description of dark energy. For  models $I_{\beta=0}$ (gray) and $I_{\alpha=0}$ (blue) the function $Om$ is increasing what means that the best model for dark energy is provided by a phantom. Remaining models have decreasing $Om$ functions and therefore they are of quintessence type. However, one can notice that
   both $I$ and $I_{\beta=0}$ models have a long period of being of cosmological constant type.
    The model parameters were fitted using SN, $H_z$, BAO and CMB data (Table \ref{zestawienie}, No 7--12).}
   \label{diagnostic_om_w_a}
\end{figure}

In this paper we have studied  and confronted against astrophysical data as well as against the standard $\Lambda$CDM
model five cosmological models obtained from two solutions presented in Sections 4, 5.  For a deeper examination
and  comparison of properties of our models we have used various cosmological parameters as: deceleration parameter
$q(a)$, effective equation of state $w_{eff}(a)$, JERK $j(a)$, SNAP $s(a)$ (see e.g. \cite{visser,jerk}).
For a later convenience we shall express them in terms of the effective potential $V(a)$
\begin{equation}\label{deceleration}
    q(t)=-\frac{1}{a}\frac{d^2a}{dt^2}\left[\frac{1}{a}\frac{da}{dt}\right]^{-2},\hspace{0.4cm}\Leftrightarrow\hspace{0.4cm}
    q(a)=-\frac{aV^\prime(a)}{2V(a)} \ .
\end{equation}

\begin{equation}\label{w_eff}
    w_{eff}(a)=\frac{1}{3}[2q(a) - 1]\ .
\end{equation}

\begin{equation}\label{jerk}
    j(t)=+\frac{1}{a}\frac{d^3a}{dt^3}\left[\frac{1}{a}\frac{da}{dt}\right]^{-3},\hspace{0.4cm}\Leftrightarrow\hspace{0.4cm}
    j(a)=\frac{a^2V^{\prime\prime}(a)}{2V(a)} \ .
\end{equation}

\begin{equation}\label{snap}
    s(t)=+\frac{1}{a}\frac{d^4a}{dt^4}\left[\frac{1}{a}\frac{da}{dt}\right]^{-4},\hspace{0.4cm}\Leftrightarrow\hspace{0.4cm}
    s(a)=\frac{a^3V^{\prime\prime\prime}(a)}{2V(a)}+
    \frac{a^3V^{\prime\prime}(a)V^\prime(a)}{4V^2(a)}\ .
\end{equation}
where $V^\prime(a)=\frac{dV}{da}$. Therefore, explicit analytic form of the potential function is also helpful for
determination of these diagnostics. Below we shall summarize the results of our analysis, for details see fig.
\ref{cmb_deceleration}, \ref{snap_jerk}, \ref{diagnostic_om_w_a}.

Deceleration parameter $q(a)$ for $\Lambda$CDM, in contrast to other models, goes asymptotically to $-1$.
However, the present day values of $q(a)$ for  new models are not much different, particulary
$q_{I_{\beta=0}}(a=1)<q_I(a=1)<q_{\Lambda CDM}(a=1)<q_{II}(a=1)$. Only for two models with  $\alpha=0$ the present day
values o $q(a=1)$ are visibly greater (see fig. \ref{cmb_deceleration} and table \ref{zestawienie}). One notices
that the values of $q(a)$ are almost the same at the point $a\approx 0.75$ for all of the models except
the ones with $\alpha=0$. At the beginning $q(a=0)\approx 0.5$  for all models except $II$ for which
one has $q_{II}(0) = -0.02$ instead.
It means a matter dominated epoch at the beginning of cosmic evolution (cf. below for plots of $w_{eff}(a)$).

Very similar situation is for effective equation of state  parameter since it is algebraically related to $q(a)$ (cf. (\ref{w_eff})). For $\Lambda$CDM model $\lim_{a \to \infty}w_{eff}(a)=-1$ and
$w_{eff,I_{\beta=0}}(a=1)<w_{eff,I}(a=1)<w_{eff,\Lambda CDM}(a=1)<w_{eff,II}(a=1)<w_{eff,I_{\alpha=0}}(a=1)
<w_{eff,II_{\alpha=0}}(a=1)$ (see bottom fig. \ref{cmb_deceleration}). At $a=0$ there is
$w_{eff,I_{\beta=0}}\approx w_{eff,I}\approx w_{eff,I_{\alpha=0}}\approx w_{eff,II_{\alpha=0}}
\approx w_{eff,\Lambda CDM}= 0$. This explicitly indicates early matter dominated epoch noticed already before
for these models.

As it can be seen on the fig. \ref{snap_jerk}, that JERK parameter  $j(a)$ for $\Lambda$CDM model is
constant $j_{\Lambda CDM}=1$ and for $a<0.5$ models $I_{\alpha=0}$, $I$ and $I_{\beta=0}$
approximate this value. Similarly,   $j_{II_{\alpha=0}}\approx 0.2$  is also constant.
From the figure one can see that different models predict different present-day values of JERK and SNAP.
Unfortunately estimations of these parameters are beyond our present observational possibilities. However some recent analysis support the current values of jerk bigger than 2 \cite{jerk}.

In order to better understand properties of our models we shall use  $Om$ diagnostic from \cite{Sahni}:
\begin{equation}\label{diagnostic_om}
     Om(z) \equiv \frac{H^2(z)/H^2_0-1}{(1+z)^{3}-1},\hspace{0.4cm}
     z>0\ .
\end{equation}
which is designed to answer the question of nature of dark energy. For $\Lambda$CDM model we have
$Om(z)=constant$  (see black line on fig. \ref{diagnostic_om_w_a}) which means that cosmological constant $\Lambda$
is the best description of dark energy. For the model $I_{\beta=0}$ the function $Om$ is increasing what means that
the best model for dark energy is provided by a phantom.
Remaining models have decreasing  $Om$ function and therefore they are of quintessence type. However, one can notice
that two models $I$ and $I_{\beta=0}$  have a long period of being of cosmological constant type.

\section{Summary and conclusions}

In this paper we have found, and then analyzed by fitting to experimental data, two families of cosmological models
based on two different solutions of Palatini modified gravity equipped with 
non-minimal curvature coupling to (free) scalar dilaton-like field. We assume Cosmological Principle to hold and standard
spatially flat FRW metric with dust matter as a source.
Our analysis reveals how rich is class of cosmological models offered
by modified gravity. Beside of two main models denoted respectively as $I$ and $II$ we have
investigated three special cases $I_{\beta=0}$, $I_{\alpha=0}$ and $II_{\alpha=0}$ corresponding to reduced
lagrangian functions. Particularly, as a by-product, we have employed quadratic a la' Starobinsky gravity,
$I_{\beta=0}$, for  description of the cosmic acceleration.
 All models labeled by $I$ correspond to the solution (\ref{standard})
which is the same solution as in Einstein gravity provided with FRW metric and dust matter. In the case
of standard gravity this solution leads, after solving Friedmann equation, to decelerating expansion.
Modification of gravitational Lagrangian (e.g. by adding cosmological constant) modifies Friedmann equation and thence
give rise to accelerating expansion. In contrast, models labeled by $II$ are based on the solution (\ref{news}) which has no
correspondence to standard gravity. Therefore the limit $\beta\mapsto 0$ cannot be performed.
Having  estimated models parameters by using recent astrophysical data one has
demonstrated that different models become in a good agreement with the data. We have also performed various comparative
analysis against $\Lambda$CDM concordance model which is commonly accepted as a best fitted model.

The first model $I$, in spite of much more complicated Friedmann equation, qualitatively and quantitatively
mimics the $\Lambda$CDM one ($\Omega_{0\Lambda}=0.77$) from the very beginning  of cosmic evolution (Big Bang)
until the recent time. As it can be seen, on various plots, effective potential, deceleration and effective equation
of state parameters (fig. \ref{potencjal_ii},
\ref{cmb_deceleration}) are almost the same until the present time ($a=1$). Differences appear for diagnostics
employing third and forth order derivatives of the scale factor (fig \ref{snap_jerk}). These properties would
be helpful for future discrimination between models. Both models exhibit existence of the initial singularity, but
in contrast to $\Lambda$CDM, our model  predicts the (final) finite-size sudden (finite-time) singularity at the point $a=2.357$.
The effective potential for
our model has one maximum at the point $a=0.552, (z=0.811)$ which corresponds to saddle critical point at the
phase portrait (see fig \ref{cmb_portret_fazowy_ii}) and representing Einstein's static solution in full analogy
with $\Lambda$CDM. For the present time the deceleration parameter $q=-0.814$ and $w_{0, eff}=-0.876$ are within
the expected estimations. The Hubble diagram for this case is practically indistinguishable from $\Lambda$CDM diagram,
cf. Fig. \ref{hubble}. Moreover, the estimation with CMB data added does not provide any essential
(qualitative nor quantitative) changes to this model.

At the fig. \ref{omegaM} are shown posterior probability density functions for $\Omega_{0,m}$ parameter
for all of our models. It's clearly seen that density function for model I almost entirely overlaps such
function for $\Lambda$CDM model (see also table \ref{zestawienie} for the corresponding best fit values of
$\Omega_{0,m}$). One can also see that inclusion to our estimations CMB data makes possible proper
estimation of $\Omega_{0,m}$ for the following cases: $I_{\beta=0}$, $II_{\alpha=0}$ and II. In particular,
for models II amount of $\Omega_{0,m}$ is one order less than for $\Lambda$CDM and it has the order of
barionic matter.

The second our model gives rise to permanently accelerating Universe without initial singularities:
Big Bang scenario is replaced by Big Bounce. The potential plot
has no maximum and there is no saddle point at the phase portrait (fig \ref{cmb_portret_fazowy_ii}). The lack of the
initial singularity is the main advantage of this model in comparison with $\Lambda$CDM model. The coincidence
with  $\Lambda$CDM effective potential holds for $a\in (0.6, 1.2)$. The acceleration parameter is almost linear for the
scale factor near zero.  The potential function indicates big slope for the scale factor near zero what can be
interpreted as inflationary era.  Effective equation of state at the beginning is close to
$-\frac{1}{3}$ (fig \ref{cmb_deceleration}), which corresponds to domination of the spatial curvature at the
initial phase of Universe's evolution. For the present time the deceleration parameter $q_0=-0.529$ and
effective equation of state $w_{0, eff}=-0.686$. Generic finite time finite size singularity has appear at $a=2.343$.
Again, on the level of Hubble diagram, one cannot distinguish it from $\Lambda$CDM model.

\begin{figure}[h!!!!]
\centering
   \includegraphics[width=0.45\textwidth]{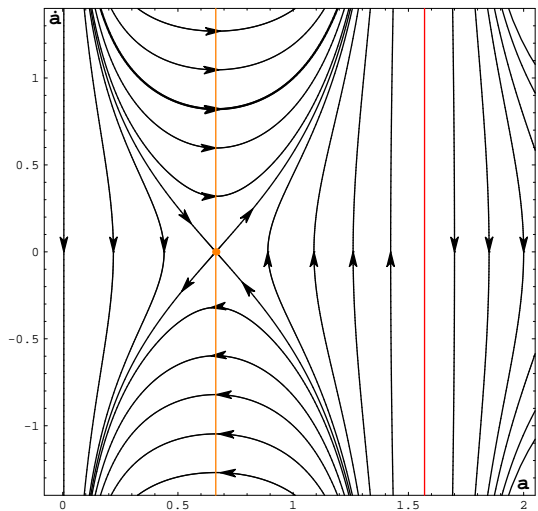}\\
   \caption{The phase portrait for quadratic gravity model $I_{\beta=0}$ on the plain $(a, \dot a)$ -- estimations including CMB data (see fig. (\ref{potencjal_ob_0_ii})). One can observe Big Bang singularity, saddle  point   corresponding to static solution as well as behaviour of the system near sudden singularity (red vertical line). Bolded trajectory of the flat model is determined by energy level $E=0$.}\label{portret_fazowy_ob_0}.
\end{figure}

In this work we have also examined the significance of the Starobinsky's term. It appears, and can be seen on the
plots of the potentials (fig \ref{potencjal_oa_0_ii}, \ref{potencjal_oa_0}) and on the Hubble's diagram
(fig \ref{hubble}), that without this term the remaining parameters of the model cannot be properly fitted.
Quadratic gravity along well qualitatively mimics $\Lambda$CDM model till the present time (see fig. (\ref{potencjal_ob_0_ii}, \ref{portret_fazowy_ob_0})). However combining it with the non-minimally coupled scalar field provides fine tuning and much better adjustment to the concordance model. It is also to be observed that the parameter $\alpha$ is negative provided that $\beta\neq 0$. In contrast for $I_{\beta=0}$ one has $\alpha>0$ which provides the so-called Chameleon effect \cite{chameleon}. We have no at the moment good explanation for this and the problem will be studded in our future work.

The Bayesian framework of model selection has been used for comparison theoretical models with
the concordance $\Lambda$CDM one. We investigated Bayes Factor to show which model is the best (most probable)
in the light of the astronomical data. We have found that while some special cases of theoretical models becomes
a weak evidence in favor of it over $\Lambda$CDM if observations from current epoch are used (SN, H(z), BAO data).
Inclusion of the information coming from early epochs (CMB data ) changes this situation because we obtain strong
evidence in favor of Standard Cosmological Model.

Our investigations  here have aimed to distinguish the favorable model by cosmography of the FLRW background metric in the sample of theoretical  models. Because of plenitude of dynamical scenarios, an introductory selection of sample of theoretical  models was necessary and it accounts in final Bayesian inference. In the Bayesian framework adding of new observations is natural for  improving models parameters. It means that the effects of cosmological perturbations in this class of models have not been considered here and this important task is postponed for future investigations. This will allow to enlarge the discriminatory tools for further analysis. For example, a sound speed of the fluctuations for  the quadratic gravity model $I_{\beta=0}$ as calculated in \cite{Koivisto2} is
\begin{equation}\label{z}
c_s^2=\frac{\Omega_{0,\alpha}}{\Omega_{0,\alpha}-a^3}
\end{equation}
which in our case $\Omega_{0,\alpha}\sim 4$ yields a superluminal value $c_s^2>1$. Thus such a model should be, in principle, rejected.
The dynamics of cosmological perturbations in extended gravity models, including Palatini formulation,
have been studied in number of papers (see e.g. \cite{Z1} and references therein) both for a matter density as well as for the background metric. However the contributions coming from non-minimal dilaton-curvature couplings one deals here have not been developed yet (cf. \cite{Z2} in the context of other scalar-tensor theories) and will be a subject of our future investigations.

\acknowledgments
A.B. gratefully acknowledges interesting discussions with G. Allemandi, S. Capozziello, M. Francaviglia and S. Odintsov
during a preliminary stage of this project. A.B. is supported by the Polish NCN grant 2011/B/S12/03354.

\FloatBarrier

\newpage

\begin{sidewaystable}
\begin{center}
\tiny\begin{tabular}{|r|c|c|c|c||c|c||c|c|}
  \hline
  \multicolumn{9}{|c|}{models I: equation (\ref{H_H0_2rozw_i}) - the parameters estimated without CMB data}\\
  \hline
  \multicolumn{2}{|c|}{}&$\Omega_{0,\alpha}$&$\Omega_{0,\beta}$&$\delta$&\multicolumn{2}{|c||}{$\Omega_{0,m}$}&$q_0$&$w_{eff,0}$\\
  \hline
  1&$\alpha=0$,\, $\Omega_{0,\beta} \in <-10,10>$,\, $\delta \in (0,1>$& - & $4.802\pm2.716 (7.007)$ & $0.268^{+0.023}_{-0.005} (0.295)$&\multicolumn{2}{|c||}{$0.02\pm 0.01 (0.01)$}&$-0.043$&$-0.362$\\
  2&$\Omega_{0,\alpha} \in <0,40>$,\, $\beta=0$& $4.401\pm0.079 (4.393)$ & $-$ & $-$ &\multicolumn{2}{|c||}{$0.37^{+0.01}_{-0.02}(0.37)$}&$-1.004$&$-1.003$\\
  3&$\Omega_{0,\alpha} \in <-30,0>$,\, $\Omega_{0,\beta} \in<-10,10>$,\, $\delta \in (0,1>$& $-18.031^{+3.911}_{-11.969}(-6.210)$ & $5.678^{+4.322}_{-1.489}(2.190)$ & $0.238^{+0.075}_{-0.010}(0.229)$ &\multicolumn{2}{|c||}{$0.25\pm0.03(0.23)$}&$-0.814$&$-0.876$\\
  \hline
  \multicolumn{9}{|c|}{models II: equation (\ref{H_H0_omegac}) - the parameters estimated without CMB data}\\
  \hline
  \multicolumn{2}{|c|}{}&$\Omega_{0,\alpha}$&$\Omega_{0,c}$&$\delta$&$\Omega_{0,\beta}$&$\Omega_{0,m}$&$q_0$&$w_{eff,0}$\\
  \hline
  4&$\alpha=0$,\, $\Omega_{0,c} \in <-1,10>$,\, $\delta \in (0,1)$& - &$2.572^{+0.613}_{-0.395}(0.651)$&$0.997\pm0.004(1.000)$&$0.254\pm0.017(0.250)$&$0.652\pm0.239(0.163)$&$-0.242$&$-0.495$\\
  5&$\Omega_{0,\alpha} \in <-60,-10>$,\, $\Omega_{0,c} \in<-1,5>$,\, $\delta \in (0,1)$& $-44.686^{+5.016}_{-15.314}(-57.870)$ & $0.715^{+0.196}_{-0.393}(0.232)$ & $0.598^{+0.008}_{-0.011}(0.560)$ & $0.009\pm0.005 (0.003)$ & $0.05^{+0.05}_{-0.04}(0.001)$&$-0.529$&$-0.686$\\
  \hline
   \multicolumn{9}{|c|}{model $\Lambda$CDM: $H^2/H_0^2= 1-\Omega_{0,m}+\Omega_{0,m}(1+z)^3$ - the parameter estimated without CMB data}\\
  \hline
  \multicolumn{2}{|c|}{}& \multicolumn{5}{|c||}{$\Omega_{0,m}$} &$q_0$&$w_{eff,0}$\\
  \hline
  6&$\Omega_{0,m} \in <0,1>$ & \multicolumn{5}{|c||}{$0.258^{+0.002}_{-0.002}(0.258)$} &$-0.613$&$-0.742$\\
  \hline
  \hline
  \multicolumn{9}{|c|}{models I: equation (\ref{H_H0_2rozw_i}) - the parameters estimated with CMB data added}\\
  \hline
  \multicolumn{2}{|c|}{}&$\Omega_{0,\alpha}$&$\Omega_{0,\beta}$&$\delta$&\multicolumn{2}{|c||}{$\Omega_{0,m}$}&$q_0$&$w_{eff,0}$\\
  \hline
  7&$\alpha=0$,\, $\Omega_{0,\beta} \in <-10,10>$,\, $\delta \in (0,1>$& $-$ & $0.0016\pm0.0007(0.0004)$ & $0.0016 \pm0.0007(0.0004)$ &\multicolumn{2}{|c||}{$0.415^{+0.029}_{-0.029}(0.409)$}& $0.355$ & $-0.097$\\
  8&$\Omega_{0,\alpha} \in <0,40>$,\, $\beta=0$& $3.854\pm 0.103 (3.843)$ & $-$ & $-$ &\multicolumn{2}{|c||}{$0.319^{+0.010}_{-0.010}(0.319)$}& $-1.236$ & $-1.157$\\
  9&$\Omega_{0,\alpha} \in <-30,0>$,\, $\Omega_{0,\beta} \in<-10,10>$,\, $\delta \in (0,1>$& $-20.824 ^{+7.144}_{-6.815} (-6.483)$ & $4.237^{+1.648}_{-1.749} (1.076)$ & $0.210^{+0.018}_{-0.019} (0.172) $ &\multicolumn{2}{|c||}{$0.266\pm0.011(0.268)$}&$-0.844$&$-0.896$\\
  \hline
  \multicolumn{9}{|c|}{models II: equation (\ref{H_H0_omegac}) - the parameters estimated with CMB data added}\\
  \hline
   \multicolumn{2}{|c|}{}&$\Omega_{0,\alpha}$&$\Omega_{0,c}$&$\delta$&$\Omega_{0,\beta}$&$\Omega_{0,m}$&$q_0$&$w_{eff,0}$\\
  \hline
  10&$\alpha=0$,\, $\Omega_{0,c} \in <-1,10>$,\, $\delta \in (0,1)$& - & $0.047\pm0.005 (0.046)$ & $0.605^{+0.019}_{-0.021} (0.604)$ & $1.989^{+0.020}_{-0.017} (1.991)$ & $0.093^{+0.004}_{-0.004} (0.093)$& $-0.029$ & $-0.352$\\
  11&$\Omega_{0,\alpha} \in <-60,-10>$,\, $\Omega_{0,c} \in<-1,5>$,\, $\delta \in (0,1)$& $-56.342^{+3.102}_{-2.971} (-59.887)$ & $1.905^{+0.230}_{-0.227} (2.000)$ & $0.615^{+0.002}_{-0.003}(0.613)$ & $0.012^{+0.001}_{-0.001}(0.012)$ & $0.023^{+0.001}_{-0.001}(0.023)$&$-0.453$&$-0.635$\\
  \hline
 \multicolumn{9}{|c|}{model $\Lambda$CDM: $H^2/H_0^2= 1-\Omega_{0,m}+\Omega_{0,m}(1+z)^3$ - the parameter estimated with CMB data added}\\
  \hline
  \multicolumn{2}{|c|}{}& \multicolumn{5}{|c||}{$\Omega_{0,m}$} &$q_0$&$w_{eff,0}$\\
  \hline
  12&$\Omega_{0,m} \in <0,1>$ & \multicolumn{5}{|c||}{$ 0.262^{+0.011}_{-0.012}(0.262)$}&$-0.608$&$-0.738$\\
  \hline
  \hline
\end{tabular}
\caption[]{The values of estimated parameters (mean of the marginalized posterior probabilities and $68 \%$ credible intervals or sample square roots of variance, together with mode of the joined posterior probabilities, shown in brackets) for all discussed models. Model $I_{\alpha=0}$ corresponds to rows No 1, 7;
model $I_{\beta=0}$: No 2, 8; model $I$: No 3, 9; model $II_{\alpha=0}$: No  4, 10; model $II$: No 5, 11;
$\Lambda$CDM: No 6, 12. Computations were made using $Union2+H_z+BAO$ data. We compare estimations without CMB data
(top part of the table) with the one employing CMB data (bottom part).}
\label{zestawienie}
\end{center}
\end{sidewaystable}


\end{document}